\documentclass[pdflatex,sn-mathphys-num]{sn-jnl}

\usepackage{graphicx}
\usepackage{amsmath,amssymb,amsfonts}
\usepackage{bm}
\usepackage{xcolor}
\usepackage{booktabs}

\begin{document}

\title[Stored energy and mechanical load in deformable dielectrics]{Stored energy and mechanical load in deformable dielectrics: What the macroscopic fields determine}

\author*[1]{\fnm{Bela} \sur{Schulte Westhoff}}\email{schultewesthoff@tu-berlin.de}

\affil*[1]{\orgname{Technische Universit{\"a}t Berlin}, \orgaddress{\postcode{10587} \city{Berlin}, \country{Germany}}}

\abstract{The mechanical load that an electric field exerts inside a deformable dielectric has been debated since Kelvin derived it from forces on dipoles and Helmholtz from an energy variation. Current electroelastic models follow the energy route, with the stored energy and its deformation dependence as constitutive input. Using a classical model of ions, electrons, and their fields and spatial averaging, we examine what this energy represents and how much of it the macroscopic fields determine. The macroscopic fields carry only their own energy. In their balance, $\mathbf J_{\rm total}\cdot\mathbf E$ is power leaving them; they do not show where it goes. In the model this power feeds carrier motion and field structure that averaging removes. For linear reversible response its time integral is $\tfrac12\mathbf E\cdot\mathbf P$: $\tfrac12\mathbf E\cdot\mathbf D$ is the macroscopic-field energy $\tfrac12\varepsilon_0|\mathbf E|^2$ plus recoverable energy of this structure. Thermal, kinetic, and elastic energies follow likewise, giving one energy map across thermodynamics, mechanics, and electrodynamics. Under deformation the same exchange supplies mechanical work; the fields fix the total, not its division between work and stored energy. We formulate a sufficiency criterion: a macroscopic description determines the load when all microstructures consistent with it give the same virtual work for the deformations of interest. For linear, nonlinear, and hysteretic response we specify the material information required; the linear case recovers the Helmholtz stress. Two dipole arrays with identical macroscopic fields, previously found to differ in boundary traction, show that sufficiency can fail; the criterion separates cases where constitutive data suffice from those needing interface or microstructural detail.}


\maketitle

\section{Introduction}
\label{sec:introduction}

Predicting how a dielectric deforms in an electric field requires the bulk force density and boundary traction that the field exerts on the material. Several expressions for this load have long coexisted~\cite{BobbioElectrodynamicsMaterials}. The Kelvin force density follows from the forces on polarized matter treated as dipoles, whereas the Korteweg--Helmholtz force density follows from the variation of a stored energy and contains an electrostrictive term~\cite{Melcher1981,Stratton1941,LandauLifshitzVol8}. The Lorentz and Einstein--Laub expressions give the same total force on a body while distributing it differently in its interior~\cite{BarnettLoudon2006ForceDielectricMedium,MansuripurZakharianWright2013}. The disagreement thus concerns the local distribution of the load, and agreement in total force need not determine the distribution that drives deformation~\cite{HausMelcher1989,Markel2023ElectrostaticForceDensity,ParkChoiParkWangPark2020}. Comparisons of force models with measured total forces and local deformations show that the models differ in the deformation they predict~\cite{Reich2017,ReichRickertMueller2018,Rickert2023}.

These expressions arise along two routes. The force route starts from the forces on microscopic constituents and asks how they load the continuum. The energy route obtains the load from the change of a stored energy under a virtual deformation. A recent analysis of the force route showed, in controlled model comparisons, that the macroscopic fields need not determine the local load~\cite{SchulteWesthoffForceIndeterminacy2026}. Bound dipoles and individually anchored charges in the same applied field load the material differently, and two dipole arrangements with the same polarization and macroscopic field produce different boundary tractions. The missing information concerns how forces on the charges are transmitted to the material and which microscopic correlations averaging removes. Here we examine the energy route.

The energy route underlies established electroelasticity, which obtains mechanical loads from variations of a stored-energy function under prescribed electrical conditions~\cite{ToupinElasticDielectric,EringenMaugin1990,DorfmannOgden2005NonlinearElectroelasticity,McMeekingLandis2005StoredEnergyDeformableDielectrics,SuoZhaoGreene2008NonlinearFieldTheory,ZhaoSuo2008}. These methods require an appropriate thermodynamic potential~\cite{Liu2013EnergyFormulationsElectrostaticsContinuumMedia} and, for dissipative materials, a description of stored energy and internal state~\cite{MieheRosato2011RateDependentFerroelectricity,FooEtAl2012DissipativeDielectricElastomers}. The stored energy and its dependence on deformation therefore enter as constitutive input.

What the stored energy represents has been examined from the microscopic side. Earlier work links material energy and heat flow to microscopic motion and interactions~\cite{DeGrootSuttorp1972}, and describes how polarization couples to thermal and mechanical work~\cite{Tiersten1971Thermoelectroelasticity}. Other studies examine how energy is divided and exchanged between fields and matter~\cite{RichterFlorianHenneberger2008Poynting,Saldanha2011DivisionEnergyMomentum,KinslerFavaroMcCall2009FourPoyntingTheorems}. For discrete dipoles, Yaghjian derived from the microscopic Maxwell equations that the macroscopic polarization work equals the work done on the bound carriers plus the change of a local field energy: that of the difference between the fields of the dipoles and the fields of the continuous polarization replacing them~\cite{Yaghjian2016ClassicalPowerEnergy}. There the work on the carriers is kept as mechanical energy of the dipole model, and the aim is a nonnegative macroscopic energy density. For periodic composites, cell-averaged quadratic expressions of the microscopic fields, the stored energy among them, can be related exactly to the macroscopic fields and the effective dielectric function~\cite{Silveirinha2009PoyntingStoredEnergy}. Both results concern energy and power in an undeformed medium.

On the side of the load, microscopic calculations have shown that macroscopic parameters do not fix its local distribution. Two-dimensional lattices of dielectric cylinders with the same effective permittivity have electrostrictive tensors and body force densities that depend on the lattice symmetry, although the total force is the same~\cite{SunWangNgZhouChan2015}; the electrostrictive terms are there understood as compensating for field information lost in the effective-medium average. The stress at a boundary between such media is indeterminate if only the macroscopic parameters are specified~\cite{WangNgXiaoChan2016}. For a model ionic crystal, the Helmholtz force density reproduces the computed strain, and it requires an electrostriction coefficient that does not appear directly in Maxwell's equations or in the constitutive relations~\cite{Markel2023ElectrostaticForceDensity}. In the rigorous continuum limit of magnetic dipoles on a lattice, the force between two parts of a body contains a local surface term that reflects the lattice~\cite{Schloemerkemper2005ContinuumLimitMagneticForce,PopovicPraetoriusSchloemerkemper2007}. Homogenization constructs stored-energy functions from a specified microstructure~\cite{PonteCastanedaSiboni2012Homogenization}.

To our knowledge, these two lines of work have not been linked through the stored energy. The first identifies energy in field structure that the macroscopic fields do not carry. The second shows that the local load depends on microstructure and expresses this through force densities and electrostriction coefficients. Here we connect them: within one classical model, what the macroscopic fields leave undetermined in the load is the change under deformation of the field energy that averaging removes. We state the resulting condition for a macroscopic description to suffice as a virtual-work criterion.

We start from the classical model of Ref.~\cite{SchulteWesthoffForceIndeterminacy2026}: stable ions, electrons, and their electromagnetic fields. In its energy balance, the two active energy stores are particle kinetic energy and microscopic field energy. The organization of carrier motion into three roles and the spatial averaging that gives the macroscopic description follow that reference; we call the spatially averaged electromagnetic fields the \emph{retained fields}. Three results follow.

First, we state what the retained fields record about energy. They form a closed Maxwell system that carries no information about the structure removed by averaging, and they carry their own energy and its flux. Their balance has a single exchange term, the product $\mathbf J_{\rm total}\cdot\mathbf E$ of total current and field: the power leaving them. Because the removed structure is confined to matter, they also determine the work supplied by the external charging source, up to corrections from unresolved surface layers (Sec.~\ref{subsec:external-power-conductor-charging}). They do not show where the exchanged energy goes. For a dielectric polarized from one state to another, $\int\mathbf E\cdot d\mathbf P$ is therefore the energy that has left the retained fields, not the energy of a polarization field.

Second, we identify what lies beyond the exchange. The product $\mathbf J_{\rm total}\cdot\mathbf E$ is not the averaged work on the carriers: part of it feeds field structure that averaging removes, as found for discrete dipoles in Ref.~\cite{Yaghjian2016ClassicalPowerEnergy}. In a model with two stores and workless constraints, the recoverable quasistatic polarization work is entirely energy of this structure. This identification is specific to the model; the exchange itself and the sufficiency criterion below do not depend on it. For linear reversible response with temperature-independent permittivity at fixed deformation, the familiar density $\tfrac12\mathbf E\cdot\mathbf D$ thus combines the retained-field energy with the increment $\tfrac12\mathbf E\cdot\mathbf P$ of what we call \emph{binding energy}: the recoverable energy stored in this unresolved field structure. For nonlinear and hysteretic response the transfer history is still known, and we identify the material information that determines its recoverable part. The same construction traces thermal, kinetic, and elastic energies to the two microscopic stores, giving a common energy map for thermodynamics, continuum mechanics, and macroscopic electrodynamics.

Third, we carry this to the load and formulate a sufficiency criterion. Under deformation the same exchange also supplies the mechanical work, and the retained fields fix the total but not its division between mechanical work and the change of binding energy. In the linear case the missing information is the strain dependence of the permittivity in the Helmholtz stress~\cite{LandauLifshitzVol8,Melcher1981}, whose dependence on microstructure is established~\cite{SunWangNgZhouChan2015,WangNgXiaoChan2016,Markel2023ElectrostaticForceDensity}. A macroscopic description determines the load of interest when all microscopic realizations consistent with its fields and supplied material data give the same virtual work, to the required accuracy, for the relevant virtual deformations. Two dipole arrangements, assumed to share the same retained fields throughout a deformation, illustrate that this condition can fail: their binding-energy variations and boundary work differ. The criterion separates cases in which constitutive data at the existing resolution complete the prediction from those that need further spatial or interface information. The energy route delivers the load in the components selected by the virtual deformations, at a resolution no finer than that of the retained fields and the supplied data (Sec.~\ref{subsec:tm-output-resolution}).

The analysis is classical and restricted to electric response. Intrinsic spin and its magnetic moments are excluded, and the load predictions concern quasistatic deformation. Stabilizing interactions that the model does not resolve, including quantum effects, enter as constraint forces that do no total work and store no energy. Sections~\ref{sec:microscopic-baseline}--\ref{sec:electromagnetic-field-retention} establish the model and its averaged descriptions. Section~\ref{sec:retained-complementary-energy} establishes what the retained fields record about energy, where the exchanged energy goes in the model, and the conditions for recoverable storage. Section~\ref{sec:variational-closure} establishes what the retained fields determine about the load, the sufficiency criterion, and its applications. The balances of Sec.~\ref{subsec:carrier-ownership-ledger} and the averaged Maxwell equations of Secs.~\ref{subsec:averaging-preserves-fields} and~\ref{subsec:retained-total-source-system} are standard and are stated to fix the notation. The results build on three further steps: the three roles of carrier motion (Sec.~\ref{subsec:carrier-kinematics-ownership}), the thermal and elastic energies obtained by averaging (Sec.~\ref{subsec:tm-mass-momentum-reservoirs}), and the comparison of retained and residual fields in two examples (Sec.~\ref{subsec:quasistatic-boundary-retention-examples}).

\section{Microscopic Model and Its Energy Stores}
\label{sec:microscopic-baseline}
\label{sec:Microscopic_Baseline}

We use electrons and ions, together with their electromagnetic fields, as an effective microscopic reference, and refer to electrons and ions collectively as \emph{carriers}. Our aim is to construct the energy descriptions of thermodynamics, continuum mechanics, and macroscopic electrodynamics from this common starting point. Averaging summarizes the same particles and fields through collective variables while leaving finer detail unresolved. Here we identify the microscopic energy stores and their exchanges, then organize the particle motions that enter these averaged descriptions.

Intrinsic spin and its associated magnetic moments are excluded from both the microscopic reference and its macroscopic descriptions. Their dynamics and exchanges of energy and angular momentum are left for future work.

\subsection{Microscopic Reference Model}
\label{subsec:carrier-ownership-ledger}
\label{subsec:microscopic-momentum-equation}
\label{subsec:passive-stabilizing-constraints}
\label{subsec:microscopic-energy-velocity-projection}
\label{subsec:microscopic-vacuum-field-energy-ledger}

Each carrier possesses both mass and charge. Its motion therefore contributes to mechanical momentum and electric current. We first define the carriers and their fields, then introduce their dynamics and stability constraints to establish the energy stores and exchanges of the reference model. The field and carrier balances of this subsection are standard~\cite{Jackson1999,FanoChuAdler1960}; we state them to fix the notation and the assumptions used below.

We work in the nonrelativistic regime. For species $s$ with microscopic number density $n_s$, charge $q_s$, mass $m_s$, and velocity $\mathbf v_s$, the number flux is $\boldsymbol\Gamma_s=n_s\mathbf v_s$. Charge and mass weight this same flux differently:
\begin{equation}
\begin{aligned}
    \rho_{\rm mic}
    &=
    \sum_s q_s n_s,
    &
    \mathbf j_{\rm mic}
    &=
    \sum_s q_s\boldsymbol\Gamma_s,
    \\
    \rho_m^{\rm mic}
    &=
    \sum_s m_s n_s,
    &
    \mathbf g_{\rm mech}^{\rm mic}
    &=
    \sum_s m_s\boldsymbol\Gamma_s .
\end{aligned}
    \label{eq:microscopic-charge-mass-bridge}
\end{equation}
Mass and charge belong to the same carriers but enter different balances: charge and current define the electromagnetic sources, whereas mass weights the carrier motion to give mechanical momentum.

The total carrier sources in Eq.~\eqref{eq:microscopic-charge-mass-bridge} generate the microscopic fields through Maxwell's equations~\cite{Jackson1999}:
\begin{equation}
\begin{aligned}
    \nabla\cdot\mathbf e_{\rm mic}
    &=
    \frac{\rho_{\rm mic}}{\varepsilon_0},
    &
    \nabla\times\mathbf e_{\rm mic}
    &=
    -\frac{\partial\mathbf b_{\rm mic}}{\partial t},
    \\
    \nabla\cdot\mathbf b_{\rm mic}
    &=
    0,
    &
    \nabla\times\mathbf b_{\rm mic}
    &=
    \mu_0\mathbf j_{\rm mic}
    +
    \mu_0\varepsilon_0
    \frac{\partial\mathbf e_{\rm mic}}{\partial t}.
\end{aligned}
    \label{eq:em-microscopic-maxwell-system}
\end{equation}

In this effective model, each carrier has a smooth, localized charge and mass distribution of small finite extent, giving finite self-field energy. The same regularization is used throughout.

Newton's law for the carriers, summed and written as a local momentum balance, is
\begin{equation}
    \frac{\partial\mathbf g_{\rm mech}^{\rm mic}}{\partial t}
    +
    \nabla\cdot\mathbf t_{\rm kin}^{\rm mic}
    =
    \mathbf f_{\rm L}^{\rm mic}
    +
    \mathbf f_{\rm con}^{\rm mic}.
    \label{eq:microscopic-closed-balance}
\end{equation}
Here $\mathbf t_{\rm kin}^{\rm mic}=\sum_s m_sn_s\mathbf v_s\otimes\mathbf v_s$ is the momentum flux carried by moving particles. Local momentum changes through this transport and through the Lorentz force $\mathbf f_{\rm L}^{\rm mic}$ and constraint reaction $\mathbf f_{\rm con}^{\rm mic}$.

Maxwell's equations express the Lorentz force equivalently through the microscopic fields:
\begin{equation}
\begin{aligned}
    \mathbf f_{\rm L}^{\rm mic}
    &=
    \rho_{\rm mic}\mathbf e_{\rm mic}
    +
    \mathbf j_{\rm mic}\times\mathbf b_{\rm mic}
    \\
    &=
    \nabla\cdot\mathbf t_{\rm vac}^{\rm mic}
    -
    \frac{\partial\mathbf g_{\rm vac}^{\rm mic}}{\partial t}.
\end{aligned}
    \label{eq:microscopic-field-momentum-balance}
\end{equation}
Here $\mathbf t_{\rm vac}^{\rm mic}$ is the vacuum Maxwell stress and $\mathbf g_{\rm vac}^{\rm mic}=\varepsilon_0\mathbf e_{\rm mic}\times\mathbf b_{\rm mic}$ is the field-momentum density~\cite{Jackson1999,FanoChuAdler1960}. This identity describes the same electromagnetic momentum exchange from the field side.

The constraint force density $\mathbf f_{\rm con}^{\rm mic}$ in Eq.~\eqref{eq:microscopic-closed-balance} represents the unresolved stabilization of both the finite carrier distributions and the material configurations they form. As in the classical Maxwell--Lorentz treatment of Eringen and Maugin~\cite[Secs.~2.2 and~2.7]{EringenMaugin1990}, we take stable charge assemblies as our starting point. This description encodes the role of omitted physics, including quantum effects, while allowing the carrier motions and material deformations considered below.

We obtain the carrier energy balance by taking the scalar product of each carrier's equation of motion with its velocity and summing over the carriers. We examine the Lorentz, constraint, and kinetic terms in turn.

For each carrier, the magnetic Lorentz force is perpendicular to its velocity, so the electromagnetic power density is
\begin{equation}
    \mathcal P_{\rm mic}
    =
    \sum_s
    \mathbf f_{{\rm L},s}^{\rm mic}\cdot\mathbf v_s
    =
    \mathbf j_{\rm mic}\cdot\mathbf e_{\rm mic}.
    \label{eq:microscopic-power}
\end{equation}
Maxwell's equations express the same power transfer through the microscopic field-energy density and flux. This is Poynting's theorem:
\begin{equation}
\begin{aligned}
    \mathbf j_{\rm mic}\cdot\mathbf e_{\rm mic}
    &=
    -
    \left(
        \frac{\partial u_{\rm vac}^{\rm mic}}{\partial t}
        +
        \nabla\cdot\mathbf S_{\rm vac}^{\rm mic}
    \right),
    \\
    u_{\rm vac}^{\rm mic}
    &=
    \frac{\varepsilon_0}{2}|\mathbf e_{\rm mic}|^2
    +
    \frac{1}{2\mu_0}|\mathbf b_{\rm mic}|^2,
    \\
    \mathbf S_{\rm vac}^{\rm mic}
    &=
    \frac{1}{\mu_0}
    \mathbf e_{\rm mic}\times\mathbf b_{\rm mic}.
\end{aligned}
    \label{eq:microscopic-poynting-balance}
\end{equation}

We next consider the power of the constraint forces $\mathbf f_{\rm con}^{\rm mic}$. We require their total actual power to vanish at every instant for the admitted motions:
\begin{equation}
\begin{aligned}
    \mathcal P_{\rm con}^{\rm mic}
    &\equiv
    \sum_s
    \mathbf f_{{\rm con},s}^{\rm mic}\cdot\mathbf v_s,
    \\
    \int_{\Omega_{\rm complete}}
    \mathcal P_{\rm con}^{\rm mic}(\mathbf x,t)\,dV
    &=
    0.
\end{aligned}
    \label{eq:constraint-workless-condition}
\end{equation}
Here $\Omega_{\rm complete}$ contains all constrained carriers, including their supports. The reactions may redistribute energy between carriers and locations, so their local power need not vanish. We also assume that the constraints have no independent energy storage and therefore introduce no additional active energy store.

The carrier kinetic-energy density and flux are
\begin{equation}
\begin{aligned}
    u_{\rm mat}^{\rm mic}
    &\equiv
    \frac12\sum_s m_sn_s|\mathbf v_s|^2,
    \\
    \mathbf s_{\rm mat}^{\rm mic}
    &\equiv
    \frac12\sum_s m_sn_s|\mathbf v_s|^2\mathbf v_s .
\end{aligned}
    \label{eq:microscopic-carrier-kinetic-moments}
\end{equation}
\begin{samepage}
Retaining the local constraint power, the carrier energy balance and microscopic Poynting theorem give~\cite{RichterFlorianHenneberger2008Poynting}
\begin{equation}
\begin{aligned}
    \frac{\partial u_{\rm mat}^{\rm mic}}{\partial t}
    +
    \nabla\cdot\mathbf s_{\rm mat}^{\rm mic}
    -\mathcal P_{\rm con}^{\rm mic}
    &=
    \mathbf j_{\rm mic}\cdot\mathbf e_{\rm mic}
    \\
    &=
    -
    \left(
        \frac{\partial u_{\rm vac}^{\rm mic}}{\partial t}
        +
        \nabla\cdot\mathbf S_{\rm vac}^{\rm mic}
    \right).
\end{aligned}
    \label{eq:microscopic-two-side-ledger}
\end{equation}
\end{samepage}
Under these assumptions, carrier kinetic energy and microscopic vacuum-field energy are the two active stores of the reference model. The balance in Eq.~\eqref{eq:microscopic-two-side-ledger} supplies the common energy reference for the macroscopic construction. In it, $\mathbf j_{\rm mic}\cdot\mathbf e_{\rm mic}$ is the complete exchange between the two stores. Its macroscopic counterpart $\mathbf J_{\rm total}\cdot\mathbf E$ has a different meaning, established in Sec.~\ref{subsec:isolated-dielectric-setup}: it is the exchange between the retained fields and everything that averaging removes, which includes part of the field energy.

\subsection{Three Roles of Carrier Motion}
\label{subsec:carrier-kinematics-ownership}
\label{subsec:kinematic-roles-handoff}
\label{sec:Microscopic_Aliases}
\label{subsec:notation-scope}

Before averaging, we reorganize the motion and configuration of the same carriers into three roles corresponding to the macroscopic domains. This change of variables retains the full microscopic information. The organization follows Ref.~\cite[Sec.~II\,B]{SchulteWesthoffForceIndeterminacy2026}.

For the positive material constituent ($a=i$) and electrons ($a=e$), separate each carrier velocity into a locally ordered part and a fluctuation. Resolve the ordered electron motion relative to the ordered ionic motion:
\begin{equation}
\begin{aligned}
    \mathbf v_a
    &=
    \mathbf U_a+\mathbf c_a,
    &
    \left\langle n_a\mathbf c_a\right\rangle_{\rm loc}
    &=
    \mathbf0,
    \\
    \mathbf U_i
    &=
    \mathbf U_H,
    &
    \mathbf U_e
    &=
    \mathbf U_H+\mathbf w,
    \qquad
    \mathbf w=\dot{\boldsymbol\xi}.
\end{aligned}
    \label{eq:microscopic-three-role-kinematics}
\end{equation}
The bracket $\langle\cdot\rangle_{\rm loc}$ denotes an average over a local neighborhood containing many carriers; it defines the ordered species velocity $\mathbf U_a$ before the spatial filtering of Sec.~\ref{sec:thermomechanical-receiver}. Here $\mathbf U_H=\mathbf U_i$ is the ordered ionic velocity, which we use as the reference motion. The ordered electronic velocity differs from it by $\mathbf w=\mathbf U_e-\mathbf U_i$, with $\boldsymbol\xi$ denoting the corresponding relative displacement. The fluctuations $\mathbf c_i$ and $\mathbf c_e$ remain explicitly included.

These variables anticipate the three macroscopic descriptions. Fluctuations $\mathbf c_i,\mathbf c_e$ contribute to thermal behavior, while ordered ionic motion $\mathbf U_H$ provides the reference for motion and deformation in the continuum-mechanical description. In a locally neutral material, common ordered motion carries mechanical momentum without a net convective charge current. Relative displacement and motion $(\boldsymbol\xi,\mathbf w)$ describe charge separation and relative current, underlying the material response in macroscopic electrodynamics. All three roles remain coupled through the same microscopic fields and constraints.

\section{Thermal and Mechanical Energies by Averaging}
\label{sec:thermomechanical-receiver}

We first construct the material's thermomechanical description from the common carrier--field reference. We allow collective motion, deformation, and thermal fluctuations while holding the material's ordered relative charge configuration fixed. Averaging then defines the continuum variables and identifies the microscopic contributions represented by kinetic, thermal, and elastic energies.

\subsection{Spatial Averaging and Scale Separation}
\label{subsec:tm-filtering-plateau}

Let $a_{\rm mic}(\mathbf x,t)$ denote any microscopic scalar, vector, or tensor field. A normalized local averaging kernel $K_\ell$ of spatial width $\ell$ defines
\begin{equation}
    \overline a_\ell(\mathbf x,t)
    \equiv
    \left\langle a_{\rm mic}\right\rangle_\ell
    =
    \int
        K_\ell(\mathbf x-\mathbf r)
        a_{\rm mic}(\mathbf r,t)\,d^3r .
    \label{eq:tm-spatial-averaging}
\end{equation}
Spatial averaging can also be understood as filtering by spatial frequency: slowly varying components are retained, while fine microscopic variation is suppressed. We call the averaged quantities \emph{retained} and the averaging operation \emph{retention}. The averaged field therefore represents selected content of the same microscopic state, with the remaining detail left unresolved. This separation changes the description while leaving the underlying physical system unchanged. The unresolved content can still contribute to energy and mechanical response.

In smooth bulk, the scale of microscopic carrier arrangements, $d_{\rm mic}$, is typically many orders of magnitude smaller than the smallest spatial-variation scale $L_{\rm TM}$ retained by the thermomechanical description. The averaging width $\ell_{\rm mech}$ can then be chosen such that
\begin{equation}
    d_{\rm mic}
    \ll
    \ell_{\rm mech}
    \ll
    L_{\rm TM}.
    \label{eq:tm-scale-separation-window}
\end{equation}
The precise kernel and width are immaterial when reasonable choices within this window give the same retained fields to the required tolerance. This interval is the \emph{bulk averaging plateau}. If no such plateau exists, the retained state must be enriched or resolved more finely.

\subsection{From Charged Carriers to the Mechanical Continuum}
\label{subsec:tm-host-kinematics}

We now construct the continuum description of the carriers, starting with one carrier and then combining them into two species continua.

For one carrier $p$ of mass $m_p$, charge $q_p$, and position $\mathbf r_p(t)$, averaging replaces its localized representation by a smooth profile over the averaging region. When the carrier extent is negligible compared with $\ell$, applying the averaging in Eq.~\eqref{eq:tm-spatial-averaging} gives the point-source approximation:
\begin{equation}
\begin{aligned}
    \rho_{m,p}^{\rm ret}(\mathbf x,t)
    &\simeq m_pK_\ell\!\left(\mathbf x-\mathbf r_p(t)\right),\\
    \rho_{q,p}^{\rm ret}(\mathbf x,t)
    &\simeq q_pK_\ell\!\left(\mathbf x-\mathbf r_p(t)\right).
\end{aligned}
    \label{eq:single-carrier-retained-mass-charge}
\end{equation}
The normalized kernel preserves the integrated mass and charge. Their represented spatial extent changes, while both still belong to the same carrier moving at $\mathbf v_p(t)$. Its retained momentum density is $\rho_{m,p}^{\rm ret}\mathbf v_p$.

Summing these overlapping profiles separately for ions and electrons produces two interpenetrating continua. Their mass densities and mean velocities are defined by
\begin{equation}
\begin{aligned}
    \rho_{m,a}
    &=
    \left\langle m_an_a\right\rangle_\ell,
    \\
    \rho_{m,a}\mathbf V_a
    &=
    \left\langle m_an_a\mathbf v_a\right\rangle_\ell
    \\
    &=
    \left\langle m_an_a(\mathbf U_a+\mathbf c_a)\right\rangle_\ell
    \\
    &\simeq
    \left\langle m_an_a\mathbf U_a\right\rangle_\ell,
    \qquad a\in\{i,e\}.
\end{aligned}
    \label{eq:tm-filtered-mass-momentum}
\end{equation}
Here the spatial averaging is chosen consistently with the ordered--fluctuating decomposition in Eq.~\eqref{eq:microscopic-three-role-kinematics}, so that $\langle m_an_a\mathbf c_a\rangle_\ell\simeq\mathbf0$ to the accuracy of the averaging. Thus $\mathbf V_a$ is the mass-weighted continuum representation of the ordered species motion $\mathbf U_a$.

At the retained resolution, these continuum fields compress the microscopic carrier description: individual particle positions and velocities are no longer tracked, and different microscopic arrangements can become indistinguishable in the retained state.

The two-fluid description separates common and relative motion through
\begin{equation}
    \mathbf V_i
    =
    \mathbf V_H,
    \qquad
    \mathbf V_e
    =
    \mathbf V_H+\mathbf W,
    \qquad
    \mathbf W
    =
    \mathbf V_e-\mathbf V_i .
    \label{eq:tm-retained-species-kinematics}
\end{equation}
For the common motion considered here, the overlapping ionic and electronic continua share the velocity $\mathbf V_i=\mathbf V_e=\mathbf V_H$. In the locally neutral bulk, their charge densities and convective charge currents cancel, while their masses and mechanical momenta add, defining the mechanical \emph{host}:
\begin{equation}
\begin{aligned}
    \rho_m
    &=
    \rho_{m,i}+\rho_{m,e},
    \\
    \rho_m\mathbf V_H
    &=
    \rho_{m,i}\mathbf V_i
    +
    \rho_{m,e}\mathbf V_e .
\end{aligned}
    \label{eq:tm-derived-host-fields}
\end{equation}
The host velocity defines the retained motion and deformation. In material coordinates $\mathbf X$, let $\boldsymbol\varphi(\mathbf X,t)$ be the host trajectory, so
\begin{equation}
\begin{aligned}
    \frac{\partial\boldsymbol\varphi}{\partial t}
    &=
    \mathbf V_H(\boldsymbol\varphi,t),
    &
    \mathbf u(\mathbf X,t)
    &=
    \boldsymbol\varphi(\mathbf X,t)-\mathbf X,
    \\
    \mathbf F
    &=
    \nabla_{\!X}\boldsymbol\varphi
    =
    \mathbf I+\nabla_{\!X}\mathbf u .
\end{aligned}
    \label{eq:tm-continuum-deformation}
\end{equation}
Thus $\rho_m$, $\mathbf V_H$, $\mathbf u$, and $\mathbf F$ describe the host's retained mass, motion, and deformation; unresolved non-affine motion remains in the microscopic reference model.

\subsection{Thermomechanical Energy Contributions}
\label{subsec:tm-mass-momentum-reservoirs}

The host's retained mass and velocity specify its continuum kinetic energy. Thermal and elastic descriptions compress information about unresolved carrier motion and field structure into macroscopic state variables.

Equation~\eqref{eq:tm-filtered-mass-momentum} fixes the mean momenta, while the averaged squared velocities retain the unresolved kinetic energy. For the common motion $\mathbf V_i=\mathbf V_e=\mathbf V_H$, the averaged microscopic kinetic energy splits exactly as
\begin{equation}
    \left\langle
        u_{\rm mat}^{\rm mic}
    \right\rangle_\ell
    =
    \underbrace{
        \frac12\rho_m|\mathbf V_H|^2
    }_{k_{\rm cont}}
    +
    \underbrace{
        \frac12
        \sum_{a\in\{i,e\}}
        \left\langle
            m_an_a|\mathbf c_{a,\ell}|^2
        \right\rangle_\ell
    }_{k_{\rm sub}} .
    \label{eq:tm-kinetic-decomposition}
\end{equation}
Here $\mathbf c_{a,\ell}=\mathbf v_a(\mathbf r,t)-\mathbf V_H(\mathbf x,t)$ denotes each carrier's velocity relative to the retained host velocity at the averaging point $\mathbf x$. The unresolved kinetic energy $k_{\rm sub}$ includes both carrier fluctuations and unresolved variation of the ordered motion. Under local equilibrium, its locally randomized part is represented by the thermal energy density $u_{\rm th}$ through macroscopic variables such as temperature, without tracking individual carrier velocities.

Microscopic electromagnetic field structure is likewise unresolved by the continuum variables. Even where the retained net charge and current vanish, the microscopic charge distributions of ions and electrons need not coincide, and their local fields can remain nonzero. Common deformation can change this microscopic arrangement and its field energy without introducing an additional ordered displacement of electrons relative to ions. Because the constraints store no energy and this energy persists when the material is at rest, the recoverable energy of deformation is field energy within this classical model; it is represented by elastic energy $u_{\rm el}$. Strain and an elastic constitutive law describe this response without resolving each carrier arrangement and its fields.
Unordered carrier motion also produces fluctuating microscopic fields. After local equilibration, their unresolved energy is represented by $u_{\rm rad}$ within the thermal description. The macroscopic thermal state therefore summarizes both randomized carrier kinetic energy and fluctuating field energy.

The resulting thermomechanical energy map is
\begin{equation}
\begin{array}{c|cc}
&
\substack{{\rm unordered}\\{\rm fluctuations}}
&
\substack{{\rm common\ motion}\\{\rm and\ deformation}}
\\
\hline
{\rm carrier\ kinetic}
&
u_{\rm th}
&
k_{\rm cont}
\\
{\rm vacuum\ field}
&
u_{\rm rad}
&
u_{\rm el}
\end{array}.
    \label{eq:tm-restricted-reservoir-map}
\end{equation}
Separating unordered fluctuations from common motion and deformation thus organizes the two microscopic stores into the four entries above, which averaging and collective state variables describe as thermal, continuum kinetic, and elastic energies.

\section{Averaged Electromagnetic Fields and What Averaging Removes}
\label{sec:electromagnetic-field-retention}

We now turn to the ordered relative displacement and motion of ions and electrons, and first establish how their electromagnetic fields are represented after averaging. Keeping the thermomechanical host fixed, we spatially average the microscopic fields, separating them into retained fields and unresolved residuals. The construction follows Ref.~\cite[Sec.~III\,A]{SchulteWesthoffForceIndeterminacy2026}.

\subsection{Averaging the Sources and Fields of Each Carrier}
\label{subsec:averaging-preserves-fields}

Section~\ref{subsec:tm-host-kinematics} introduced the averaged mass and charge profiles of a single ion or electron. We now follow the corresponding electromagnetic fields through the same spatial averaging, before combining the carrier contributions by Maxwell superposition.

We use the filter of Sec.~\ref{subsec:tm-filtering-plateau} with a normalized, centered kernel and a fixed bulk averaging width $\ell_{\rm EM}$ in the electromagnetic scale-separation window:
\begin{equation}
\begin{gathered}
    \int K_{\ell_{\rm EM}}(\mathbf r)\,d^3r
    =1,
    \qquad
    \int \mathbf rK_{\ell_{\rm EM}}(\mathbf r)\,d^3r
    =\mathbf0,
    \\
    d_{\rm mic}
    \ll
    \ell_{\rm EM}
    \ll
    L_{\rm EM}.
\end{gathered}
    \label{eq:scale-separation-gap}
\end{equation}
Here $d_{\rm mic}$ is the carrier or lattice scale and $L_{\rm EM}$ the smallest retained field-variation scale. At the electromagnetic retention scale, the same averaging is applied to each carrier's mass, charge, and motion and to the corresponding fields. We write $\langle\cdot\rangle\equiv\langle\cdot\rangle_{\ell_{\rm EM}}$ for this fixed filter below.

For a carrier of charge $q$, position $\mathbf r_q(t)$, and velocity $\mathbf v_q(t)$, the averaged charge and current densities are, in the same point-source approximation,
\begin{equation}
\begin{aligned}
    \rho_q
    &=
    \left\langle\rho_q^{\rm mic}\right\rangle
    \simeq
    qK_{\ell_{\rm EM}}\!\left(\mathbf x-\mathbf r_q(t)\right),
    \\
    \mathbf J_q
    &=
    \left\langle\mathbf j_q^{\rm mic}\right\rangle
    \simeq
    q\mathbf v_q(t)
    K_{\ell_{\rm EM}}\!\left(\mathbf x-\mathbf r_q(t)\right).
\end{aligned}
    \label{eq:single-carrier-retained-sources}
\end{equation}
Normalization and centering preserve total charge and charge center, and the fixed filter preserves continuity, $\partial_t\rho_q+\nabla\cdot\mathbf J_q=0$.

We now separate each carrier's microscopic fields into retained averages and residual fields containing the unresolved detail. The retained fields satisfy Maxwell's equations with the averaged charge and current densities, because Maxwell's equations are linear and the fixed bulk filter commutes with spatial and time derivatives~\cite[Sec.~6.6]{Jackson1999}. We write this separation as
\begin{equation}
\begin{aligned}
    \mathbf E_q
    &=
    \left\langle\mathbf e_q^{\rm mic}\right\rangle,
    &
    \mathbf B_q
    &=
    \left\langle\mathbf b_q^{\rm mic}\right\rangle,
    \\
    \mathbf e_q^{\rm mic}
    &=
    \mathbf E_q+\delta\mathbf e_q,
    &
    \mathbf b_q^{\rm mic}
    &=
    \mathbf B_q+\delta\mathbf b_q .
\end{aligned}
    \label{eq:single-carrier-filtered-fields}
\end{equation}
The residual sources $\delta\rho_q=\rho_q^{\rm mic}-\rho_q$ and $\delta\mathbf j_q=\mathbf j_q^{\rm mic}-\mathbf J_q$ also satisfy continuity, and $\delta\rho_q$ has zero total charge and first charge moment. Its leading charge multipoles therefore vanish. At distances $d\gg\ell_{\rm EM}$, the microscopic and retained fields agree to the required accuracy, whereas averaging appreciably changes the field within the $O(\ell_{\rm EM})$ carrier neighborhood. The residual has decaying higher-multipole tails and need not vanish exactly. For time-dependent sources, this statement applies over the retained bandwidth when $\omega_{\max}\ell_{\rm EM}/c\ll1$.

Maxwell superposition extends this construction to the complete material by summing the carrier contributions. The retained field preserves the resolved behavior, while carrier-scale source geometry and near fields enter the residual. This linearity concerns Maxwell's equations for a specified source history; the material response itself may be nonlinear.

\subsection{Dielectric and Conductor Examples}
\label{subsec:quasistatic-boundary-retention-examples}
\label{subsec:sharp-interface-retention}

We now use a polarized dielectric and a conductor to illustrate where averaging removes microscopic field structure. In each case, we compare the microscopic and retained fields while holding the thermomechanical host fixed. The comparison isolates the ordered relative charge rearrangement and its fields; the unchanged microscopic background of the host is omitted, so the quantities shown are the field-induced changes.

\begin{figure}[!t]
    \centering
    \includegraphics[width=\textwidth]{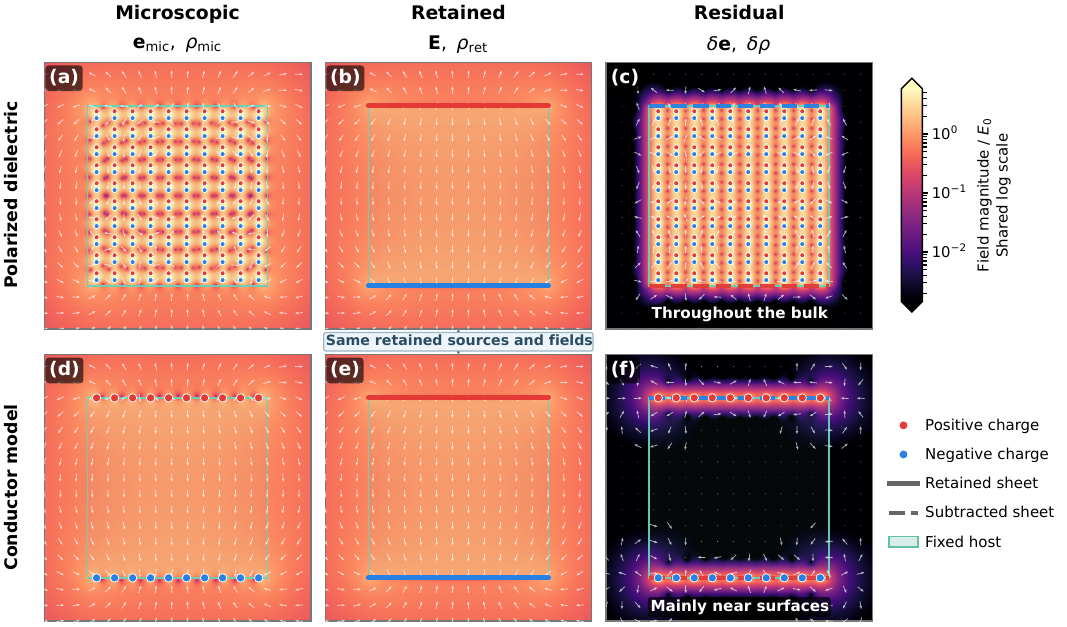}
    \caption{Conceptual illustration of identical retained sources and induced fields with different residual structure. A $10\times10$ array of bulk charge pairs (dielectric, top) and a prescribed surface-charge model (conductor, bottom) use the same retained sheets, making (b) and (e) identical. Columns show the corresponding field and charge-density decompositions, $\mathbf e_{\rm mic}=\mathbf E+\delta\mathbf e$ and $\rho_{\rm mic}=\rho_{\rm ret}+\delta\rho$; the external field is omitted. Residuals extend throughout the dielectric bulk in (c) and mainly near surfaces in (f). Color shows field magnitude on a shared logarithmic scale normalized by $E_0$, the retained field magnitude at the body center; arrows show direction. Dashed sheets represent subtracted retained charge; the green outline marks the fixed host. Both models use a common regularized two-dimensional electrostatic kernel; the lower row prescribes charges without solving conductor equilibrium. The unresolved field energy also includes cross terms with the retained field (Sec.~\ref{sec:retained-complementary-energy}).}
    \label{fig:source-field-retention-comparison}
\end{figure}

For either body, let $\rho_{\rm ret}$ and $\mathbf E$ denote the retained charge density and electric field. Defining the residuals as the differences from the microscopic quantities gives
\begin{equation}
\begin{aligned}
    \rho_{\rm mic}
    &=
    \rho_{\rm ret}
    +
    \delta\rho,
    \\
    \mathbf e_{\rm mic}
    &=
    \mathbf E
    +
    \delta\mathbf e .
\end{aligned}
    \label{eq:source-field-linear-retention-split}
\end{equation}
This decomposition is exact for the chosen retained description. The full field is $\mathbf e_{\rm mic}=\mathbf e_{\rm ext}+\mathbf e_{\rm ind}$; Fig.~\ref{fig:source-field-retention-comparison} compares the induced fields, with the common prescribed external field omitted.

\smallskip
\noindent\emph{Polarized dielectric.}
An applied field polarizes a dielectric by changing its microscopic charge organization. The array of localized charge pairs in panels (a)--(c) provides an illustrative configuration; the averaging construction applies directly to the underlying ions and electrons without requiring a particular polarization mechanism. Even where the averaged bulk charge density vanishes, microscopic charge structure and its fields remain. Averaging suppresses this detail in the retained fields, leaving residual fields throughout the bulk and near its boundary in this example.

\smallskip
\noindent\emph{Conductor.}
In the quasistatic perfect-conductor limit, mobile carriers redistribute excess charge to the surface. Its induced field cancels the applied field in the resolved neutral bulk. Panels (d)--(f) of Fig.~\ref{fig:source-field-retention-comparison} illustrate this surface organization with a prescribed charge distribution. Averaging leaves residual fields mainly near the surface, whereas the dielectric example has residual fields throughout the bulk.

Panels (b) and (e) have identical retained sources and induced fields by construction, whereas panels (c) and (f) show different residual structures. Different microscopic arrangements can therefore share the same retained source-and-field description.

\smallskip
\noindent\emph{From a filtered boundary layer to a sharp interface.}
In smooth bulk, the scale separation in Eq.~\eqref{eq:scale-separation-gap} makes the retained fields insensitive to admissible changes of the averaging kernel. At a sharp material boundary, the kernel overlaps matter and vacuum, and spatial variation extends across the averaging scale. No clear spectral gap selects a unique boundary profile: averaging produces a smooth transition layer of thickness $O(\ell_{\rm EM})$ whose shape depends on the chosen kernel.

The sharp surface sheets in panels (b) and (e) replace this smooth transition profile. To relate them to the averaging construction, first write the literally filtered charge density and field as
\begin{equation}
\begin{aligned}
    \rho_{\rm filt}
    &=
    \left\langle
        \rho_{\rm mic}
    \right\rangle,
    &
    \delta\rho_{\rm filt}
    &=
    \rho_{\rm mic}
    -
    \rho_{\rm filt},
    \\
    \mathbf E_{\rm filt}
    &=
    \left\langle
        \mathbf e_{\rm mic}
    \right\rangle,
    &
    \delta\mathbf e_{\rm filt}
    &=
    \mathbf e_{\rm mic}
    -
    \mathbf E_{\rm filt}.
\end{aligned}
    \label{eq:literal-source-field-filtering}
\end{equation}

We replace this kernel-dependent transition profile by a sharp interface carrying the same excess charge per unit area, provided the replacement reproduces the filtered fields outside the transition layer to the required accuracy~\cite{Jackson1999,HausMelcher1989}:
\begin{equation}
\begin{aligned}
    \sigma(\mathbf s)
    &=
    \int_{\rm layer}
        \left[
            \rho_{\rm filt}(\mathbf s,n)
            -
            \rho_{\rm bulk}(\mathbf s,n)\chi_{\Omega}(\mathbf s,n)
        \right]dn,
    \\
    \rho_{\rm ret}
    &=
    \rho_{\rm bulk}\chi_{\Omega}
    +
    \sigma(\mathbf s)\delta_{\partial\Omega}.
\end{aligned}
    \label{eq:generic-sharp-interface-representation}
\end{equation}
Here $\rho_{\rm filt}(\mathbf s,n)$ is the literally filtered charge density of Eq.~\eqref{eq:literal-source-field-filtering} in the layer, with outward normal coordinate $n$ and tangential position $\mathbf s$; the subtracted step profile prevents bulk charge from being counted again in $\sigma$. The indicator function $\chi_{\Omega}$ equals one inside the body $\Omega$ and zero outside, and $\delta_{\partial\Omega}$ is the surface delta distribution on its boundary. For the neutral bulk of both examples, $\rho_{\rm bulk}=0$.

The decomposition in Eq.~\eqref{eq:source-field-linear-retention-split} applies to both representations. For literal filtering, the residual contains microscopic structure suppressed by averaging. With a sharp interface, it also includes transition-layer detail omitted by the surface sheet. The residual field is generated by the residual sources and may have decaying tails beyond the carrier and interface neighborhoods. The following subsection collects Maxwell's equations for the retained bulk and surface sources.

\subsection{Maxwell's Equations for the Retained Fields}
\label{subsec:retained-total-source-system}

For the total current and magnetic field, the corresponding decomposition is
\begin{equation}
\begin{aligned}
    \mathbf j_{\rm mic}
    &=
    \mathbf J_{\rm total}+\delta\mathbf j,
    &
    \mathbf b_{\rm mic}
    &=
    \mathbf B+\delta\mathbf b .
\end{aligned}
    \label{eq:current-magnetic-retained-residual-split}
\end{equation}
In smooth bulk, the retained quantities are the filtered sources and fields. Linearity of Maxwell's equations and commutation of the fixed filter with derivatives give
\begin{equation}
\begin{aligned}
    \nabla\cdot\mathbf E
    &=
    \frac{\rho_{\rm ret}}{\varepsilon_0},
    &
    \nabla\times\mathbf E
    &=
    -\frac{\partial\mathbf B}{\partial t},
    \\
    \nabla\cdot\mathbf B
    &=
    0,
    &
    \nabla\times\mathbf B
    &=
    \mu_0\mathbf J_{\rm total}
    +
    \mu_0\varepsilon_0
    \frac{\partial\mathbf E}{\partial t}.
\end{aligned}
    \label{eq:retained-maxwell-system}
\end{equation}
At unresolved interfaces, the sharp representatives of Sec.~\ref{subsec:sharp-interface-retention} satisfy these equations distributionally, with the retained surface sources and jump conditions.

Within this reference model, the retained Maxwell system follows directly from the charge and current of ions and electrons, without an intermediate molecular model or multipole approximation. Electronic or ionic displacement and molecular reorientation enter as different organizations of the same carrier sources. The field equations therefore retain the same form across these mechanisms. For specified sources and controls, the appropriate initial and boundary data complete the electromagnetic problem; material laws supply the source evolution and energetic response.

\smallskip
\noindent\emph{Source representations.}
To connect this description with conventional macroscopic notation, we introduce polarization and magnetization as an optional representation of the retained carrier sources. Choose a free/bound split, assigning any global transport or harmonic component to $\mathbf J_{\rm free}$. The remaining current can be written as~\cite{HausMelcher1989}
\begin{equation}
\begin{aligned}
    \mathbf J_{\rm total}
    &=
    \mathbf J_{\rm free}+\mathbf J_{\rm b},
    &
    \mathbf J_{\rm b}
    &=
    \frac{\partial\mathbf P}{\partial t}
    +
    \nabla\times\mathbf M .
\end{aligned}
    \label{eq:optional-retained-current-representation}
\end{equation}
The curl term is divergence-free. If the free and total charge--current pairs each satisfy continuity, then
\begin{equation}
    \rho_{\rm ret}
    =
    \rho_{\rm free}
    -
    \nabla\cdot\mathbf P,
    \label{eq:optional-retained-charge-representation}
\end{equation}
after any time-independent reference charge is absorbed into $\rho_{\rm free}$. The representation is nonunique and changes neither the total charge and current nor the Maxwell problem.

The associated excitation fields
\begin{equation}
    \mathbf D
    =
    \varepsilon_0\mathbf E+\mathbf P,
    \qquad
    \mathbf H
    =
    \frac{\mathbf B}{\mu_0}-\mathbf M .
    \label{eq:retained-composite-fields}
\end{equation}
put Eq.~\eqref{eq:retained-maxwell-system} into its familiar excitation form~\cite{KinslerFavaroMcCall2009FourPoyntingTheorems}. The electric examples below set $\mathbf M=\mathbf0$, so $\mathbf J_{\rm total}=\mathbf J_{\rm free}+\partial_t\mathbf P$.

Polarization and magnetization can describe material organization and serve as constitutive variables. These source representations and the associated excitation fields add no independent electromagnetic degrees of freedom, and do not by themselves specify an energy store or its state dependence.

The residual fields satisfy Maxwell's equations of the same form with the residual sources $\delta\rho$ and $\delta\mathbf j$. Because the equations are linear, the retained system in Eq.~\eqref{eq:retained-maxwell-system} is closed without them: no residual quantity enters it, and its solution carries no information about the residual fields, as panels (c) and (f) of Fig.~\ref{fig:source-field-retention-comparison} show for identical retained fields. The two parts remain connected through the carriers, whose motion generates both sets of sources, and they meet again in every quantity quadratic in the fields, such as energy and stress. Field structure left unresolved by averaging can therefore still contribute to the macroscopic energy.

\section{Energy of the Retained Fields and of the Unresolved Field Structure}
\label{sec:retained-complementary-energy}
\label{sec:dielectric-examples}
\label{sec:source-alias-form}
\label{sec:host-allocation-examples}

The retained fields obey Maxwell's equations with the retained sources (Sec.~\ref{subsec:retained-total-source-system}). Their energy balance therefore contains their own energy, its flux, and one exchange term, the product $\mathbf J_{\rm total}\cdot\mathbf E$ of total current and field. This section establishes what that term means and what lies on its other side.

To isolate the response of relative charge motion, we initially exclude energy exchange with the continuum-mechanical and thermodynamic domains: the host remains stationary and rigid, and thermal energy exchange is suppressed.

We begin with conductor charging to establish when the retained fields reproduce the externally supplied work. This step comes first because the exchange inside a dielectric can be assigned only when the energy entering through the retained fields is the energy actually supplied. We then examine the exchange in a dielectric and its representation as storage. Finally, we combine these contributions into the common energy map and restore mechanical and thermal energy exchange.

\subsection{Conductor Charging: Preservation of the External Work}
\label{subsec:external-power-conductor-charging}

The quasistatic conductor of Sec.~\ref{subsec:quasistatic-boundary-retention-examples} has its appreciable charging-induced residual fields concentrated near its surface. Their energetic effect is tested through the correction to the charging power.

Consider a capacitor with perfectly conducting electrodes, charged quasistatically by an external source, with the host fixed, $\mathbf U_H=\mathbf0$. Carrier inertia, changes in carrier kinetic energy, and its transport are neglected on the charging time scale. We represent the source's electromagnetic action by an external force density $\mathbf f_{{\rm ext},s}^{\rm mic}$ on each carrier species. Its field is excluded from $(\mathbf e_{\rm mic},\mathbf b_{\rm mic})$, which contain the fields generated by the redistributed capacitor charges.

Including the external drive in Eq.~\eqref{eq:microscopic-two-side-ledger} gives
\begin{equation}
\begin{aligned}
    p_{\rm ext}^{\rm mic}
    &\equiv
    \sum_s\mathbf f_{{\rm ext},s}^{\rm mic}\cdot\mathbf v_s
    \\
    p_{\rm ext}^{\rm mic}+\mathcal P_{\rm con}^{\rm mic}
    &\simeq
    -\mathbf j_{\rm mic}\cdot\mathbf e_{\rm mic}
    \\
    &=
    \frac{\partial u_{\rm vac}^{\rm mic}}{\partial t}
    +\nabla\cdot\mathbf S_{\rm vac}^{\rm mic} .
\end{aligned}
    \label{eq:externally-driven-conductor-field-energy}
\end{equation}
In Eq.~\eqref{eq:externally-driven-conductor-field-energy}, $\partial_tu_{\rm vac}^{\rm mic}+\nabla\cdot\mathbf S_{\rm vac}^{\rm mic}$ measures local energy transfer into the included electromagnetic field: positive values indicate input and negative values output. Where this exchange vanishes, any change in local field energy is balanced by transport. Under the charging assumptions, external and constraint power together supply this local transfer. Constraint reactions can redistribute energy without providing an independent store; their contribution cancels over the complete constrained capacitor assembly, including its supports. The integrated field balance therefore determines the total external charging power.

To determine whether retention preserves this total input, insert $\mathbf e_{\rm mic}=\mathbf E+\delta\mathbf e$ and $\mathbf j_{\rm mic}=\mathbf J_{\rm total}+\delta\mathbf j$ into the microscopic power product:
\begin{equation}
\begin{aligned}
    \mathbf j_{\rm mic}\cdot\mathbf e_{\rm mic}
    &=
    \mathbf J_{\rm total}\cdot\mathbf E
    +
    r_P^{\rm mic},
    \\
    r_P^{\rm mic}
    &\equiv
    \delta\mathbf j\cdot\mathbf E
    +
    \mathbf J_{\rm total}\cdot\delta\mathbf e
    \\
    &\quad+
    \delta\mathbf j\cdot\delta\mathbf e .
\end{aligned}
    \label{eq:conductor-power-residual}
\end{equation}
The residual $r_P^{\rm mic}$ is the power contribution omitted from the retained product. In the smooth bulk of this charging model, $\delta\mathbf e\simeq\mathbf0$ and the scale gap in Eq.~\eqref{eq:scale-separation-gap} gives $\langle\delta\mathbf j\cdot\mathbf E\rangle\simeq0$. Hence
\begin{equation}
\begin{aligned}
    \left\langle
        r_P^{\rm mic}
    \right\rangle
    &\simeq
    0,
    \\
    \left\langle
        \mathbf j_{\rm mic}\cdot\mathbf e_{\rm mic}
    \right\rangle
    &\simeq
    \mathbf J_{\rm total}\cdot\mathbf E
    \qquad
    \text{in smooth bulk}.
\end{aligned}
    \label{eq:conductor-retained-bulk-power}
\end{equation}
The remaining contribution of $r_P^{\rm mic}$ is concentrated in the unresolved surface neighborhood $\mathcal L_{\partial}$ of thickness $O(\ell_{\rm EM})$. Its small thickness alone does not bound the integrated power, so we require
\begin{equation}
    \left|
    \int_{\mathcal L_{\partial}}
        r_P^{\rm mic}
    \,dV
    \right|
    \leq
    P_{\rm tol}.
    \label{eq:conductor-boundary-power-tolerance}
\end{equation}
Here $P_{\rm tol}$ is the allowed surface correction to the total charging power. In smooth bulk, define the retained external power density as $p_{\rm ext}\equiv\langle p_{\rm ext}^{\rm mic}\rangle$. The averaged charging balance and the retained Maxwell system, Eq.~\eqref{eq:retained-maxwell-system}, then give~\cite{Jackson1999,HausMelcher1989,Melcher1981}
\begin{equation}
\begin{aligned}
    p_{\rm ext}
    +\left\langle\mathcal P_{\rm con}^{\rm mic}\right\rangle
    &\simeq
    -\mathbf J_{\rm total}\cdot\mathbf E
    \\
    &=
    \frac{\partial u_{\rm vac}}{\partial t}
    +\nabla\cdot\mathbf S_{\rm vac},
    \\
    u_{\rm vac}
    &=\frac{\varepsilon_0}{2}|\mathbf E|^2
    +\frac{1}{2\mu_0}|\mathbf B|^2,
    \\
    \mathbf S_{\rm vac}
    &=\frac{1}{\mu_0}\mathbf E\times\mathbf B .
\end{aligned}
    \label{eq:retained-total-source-poynting-balance}
\end{equation}
Under the stated charging assumptions and surface-power tolerance, the integrals of the microscopic balance in Eq.~\eqref{eq:externally-driven-conductor-field-energy} and of the retained balance in Eq.~\eqref{eq:retained-total-source-poynting-balance} give the same total external charging power. The accumulated charging work is then accounted for by changes in retained-field energy and net energy transport through the assembly boundary. The retained fields therefore suffice for this energy prediction without resolving the microscopic surface fields. In a dielectric, by contrast, microscopic fields omitted by averaging can contribute appreciably to energy storage.

\subsection{Dielectric Polarization: Exchange and Storage}
\label{subsec:isolated-dielectric-setup}

We now place a neutral dielectric body $\Omega_{\rm d}$ in the capacitor before charging begins, without electrical contact with the conductors. The ionic host remains stationary and rigid, $\mathbf U_H=\mathbf V_H=\mathbf0$, and the thermal energy contributions $(u_{\rm th},u_{\rm rad})$ remain unchanged. We retain the kinetic energy of ordered electron motion relative to the host, alongside the associated electromagnetic field energies.

\smallskip
\noindent\textit{External charging work.}
With the dielectric in place, we first examine the work supplied by the source through the capacitor electrodes. The dielectric influences this work through its field at the electrodes, which retention preserves to the required accuracy away from unresolved interface layers (Sec.~\ref{sec:electromagnetic-field-retention}). In the gap between the conductors and the dielectric the retained fields are thus the microscopic fields, and the retained Poynting flux reaching the body is the actual energy flux. Under the charging and boundary-power conditions of Sec.~\ref{subsec:external-power-conductor-charging}, the retained fields therefore determine the external work supplied to the complete conductor--dielectric assembly, including its supports.

\smallskip
\noindent\textit{Microscopic energy exchange.}
We now turn to the energy exchange inside the dielectric. The capacitor field and the dielectric's own fields are both included in $(\mathbf e_{\rm mic},\mathbf b_{\rm mic})$. The separate external drive acts only on the conductor charges, so the microscopic balance inside the dielectric follows from Eq.~\eqref{eq:microscopic-two-side-ledger}:
\begin{equation}
\begin{aligned}
    \frac{\partial u_{\rm mat}^{\rm mic}}{\partial t}
    +\nabla\cdot\mathbf s_{\rm mat}^{\rm mic}
    -\mathcal P_{\rm con}^{\rm mic}
    &=\mathbf j_{\rm mic}\cdot\mathbf e_{\rm mic}
    \\
    &=-\left(
        \frac{\partial u_{\rm vac}^{\rm mic}}{\partial t}
        +\nabla\cdot\mathbf S_{\rm vac}^{\rm mic}
    \right).
\end{aligned}
    \label{eq:isolated-dielectric-microscopic-redistribution}
\end{equation}
With the ionic host fixed and thermal exchange excluded, the active kinetic response is the ordered motion of electrons relative to the ions, including its kinetic-energy storage and transport. The constraint term retains the reactions associated with the imposed restrictions and their possible local redistribution of energy; their total power vanishes over the complete constrained assembly.

We now separate this microscopic power into a retained contribution and a residual, using Eq.~\eqref{eq:conductor-power-residual}:
\begin{equation}
    \mathbf j_{\rm mic}\cdot\mathbf e_{\rm mic}
    =\mathbf J_{\rm total}\cdot\mathbf E+r_P^{\rm mic}.
    \label{eq:dielectric-local-power-residual}
\end{equation}
The residual can remain appreciable throughout the bulk, so the retained current--field product need not equal the microscopic Lorentz power. Using the microscopic and retained Maxwell equations, the same power residual can equivalently be written in field form, as in Sec.~\ref{sec:microscopic-baseline}~\cite{Yaghjian2016ClassicalPowerEnergy}:
\begin{equation*}
    r_P^{\rm mic}
    =-\left(
        \frac{\partial u_{\rm unres}^{\rm mic}}{\partial t}
        +\nabla\cdot\mathbf S_{\rm unres}^{\rm mic}
      \right).
\end{equation*}
Here $u_{\rm unres}^{\rm mic}$ and $\mathbf S_{\rm unres}^{\rm mic}$ are the unresolved field-energy density and flux, defined by
\begin{align}
    u_{\rm unres}^{\rm mic}
    &\equiv u_{\rm vac}^{\rm mic}-u_{\rm vac}
    \nonumber\\
    &=\varepsilon_0\mathbf E\cdot\delta\mathbf e
      +\frac{1}{\mu_0}\mathbf B\cdot\delta\mathbf b
      +\frac{\varepsilon_0}{2}|\delta\mathbf e|^2
      +\frac{1}{2\mu_0}|\delta\mathbf b|^2,
    \label{eq:microscopic-complementary-field-energy}
    \\
    \mathbf S_{\rm unres}^{\rm mic}
    &\equiv\mathbf S_{\rm vac}^{\rm mic}-\mathbf S_{\rm vac}
    \nonumber\\
    &=\frac{1}{\mu_0}\left(
        \mathbf E\times\delta\mathbf b
        +\delta\mathbf e\times\mathbf B
        +\delta\mathbf e\times\delta\mathbf b
      \right).
    \label{eq:microscopic-complementary-field-flux}
\end{align}
In the smooth bulk, the scale separation established in Sec.~\ref{sec:electromagnetic-field-retention} makes the averaged retained--residual cross terms negligible. Residual-field squares nevertheless survive averaging: microscopic field structure can disappear from the retained fields while still contributing to the energy. Near unresolved interfaces, spatial variation extends across the averaging scale, so the bulk scale separation no longer guarantees negligible averaged cross terms. The exact unresolved energy includes the retained--residual cross terms and the transition-layer contributions omitted by the sharp-interface representation, and need not be positive pointwise.

Using this field identity to rewrite Eq.~\eqref{eq:isolated-dielectric-microscopic-redistribution} in terms of the retained Lorentz power gives
\begin{equation}
\begin{aligned}
    \mathbf J_{\rm total}\cdot\mathbf E
    &=\frac{\partial}{\partial t}
      \left(u_{\rm mat}^{\rm mic}+u_{\rm unres}^{\rm mic}\right)
    \\
    &\quad+\nabla\cdot
      \left(\mathbf s_{\rm mat}^{\rm mic}+\mathbf S_{\rm unres}^{\rm mic}\right)
      -\mathcal P_{\rm con}^{\rm mic}.
\end{aligned}
    \label{eq:dielectric-retained-receiving-balance}
\end{equation}

\smallskip
\noindent\textit{What the retained fields record.}
Equation~\eqref{eq:dielectric-retained-receiving-balance} is exact, and it separates what the retained description contains from what averaging has removed. Its left-hand side is the exchange term of the retained Poynting balance, Eq.~\eqref{eq:retained-total-source-poynting-balance}, and is fixed by the retained fields alone. Its right-hand side contains only quantities that the retained fields do not carry: carrier kinetic energy, the energy of the field structure removed by averaging, their fluxes, and constraint power. The product $\mathbf J_{\rm total}\cdot\mathbf E$ is therefore the rate at which energy leaves the retained fields. It is not the averaged work on the carriers, because part of it feeds the unresolved field energy.

Consider the polarization of the dielectric from one state to another. The source work enters the retained fields at the conductors and is carried by the retained Poynting flux. Part of it raises the retained-field energy $u_{\rm vac}$. Where polarization current flows along $\mathbf E$, energy leaves the retained fields at the rate $\mathbf E\cdot\partial_t\mathbf P$. At the end, the retained fields have recorded three quantities: the source work, the change of their own energy, and the amount that has left, $\int\mathbf E\cdot d\mathbf P$. Only the second is a property of the field states at the beginning and end; the other two are known from the history. Behind the exchange, the carriers have been displaced against their constraints and the field structure between them has changed; the energy that left is found there or has been passed on. None of this is visible in the retained fields. At a given instant they do not distinguish a dielectric being polarized from a conductor dissipating: in both, $\mathbf J_{\rm total}\cdot\mathbf E>0$. The difference lies in whether the energy returns when the process is reversed, which is a property of the material response.

This reading closes over the body without a storage assumption. To the accuracy of the averaging, residual fields are confined to matter and its surface layers (Sec.~\ref{subsec:averaging-preserves-fields}). Integrating Eq.~\eqref{eq:dielectric-retained-receiving-balance} over a volume whose boundary lies in the vacuum around the dielectric therefore leaves, to this accuracy, no flux of carrier energy or unresolved field energy. Under the assumptions of this section, the energy that has left the retained fields inside the body equals the change of carrier kinetic energy and unresolved field energy of the body, apart from constraint power, which vanishes over the complete assembly. Where in the body this energy resides is not determined at this stage.

The two sides of Eq.~\eqref{eq:dielectric-retained-receiving-balance} also differ in status. The exchange belongs to the retained description: neither a constitutive law nor microscopic detail enters it, and the reading holds for every material response. The destination is stated as the classical reference model resolves it. A finer model would divide it differently, for instance by assigning energy to the quantum effects represented here by workless constraints, but the amount is fixed by the retained fields and would be the same.

\smallskip
\noindent\textit{Storage representation.}
We now ask how to represent the right-hand side of Eq.~\eqref{eq:dielectric-retained-receiving-balance} when only the retained fields are available and the microscopic details are no longer resolved. As for the thermomechanical host in Sec.~\ref{sec:thermomechanical-receiver}, we introduce macroscopic quantities that represent the energetic effects of the unresolved response. We first express the exchange through two powers, a power received by the unresolved field structure and a relative kinetic power:
\begin{equation}
    \mathbf J_{\rm total}\cdot\mathbf E
    \simeq\mathcal P_{\rm unres}
      +\mathcal P_{\rm rel}^{\rm kin}.
    \label{eq:dielectric-retained-receiving-power}
\end{equation}
Here $\mathcal P_{\rm unres}$ represents exchange with the unresolved electrical field contribution, while $\mathcal P_{\rm rel}^{\rm kin}$ represents the kinetic response of ordered electron motion relative to the fixed ions. These powers include the associated storage, transport, and constraint-mediated transfers. The retained fields determine their sum; a material description determines their allocation and subsequent evolution.

For the constrained, nondissipative response considered here, we describe these powers through binding and ordered electron kinetic storage. At the retained resolution and to the stated accuracy, this storage representation assumes negligible net transport of these energies across each material element's boundary and negligible net constraint power within it. Under these conditions, Eq.~\eqref{eq:dielectric-retained-receiving-power} becomes
\begin{equation}
\begin{aligned}
    \frac{\partial u_{\rm bind}}{\partial t}
    +\frac{\partial k_{\rm rel}}{\partial t}
    &\simeq\mathbf J_{\rm total}\cdot\mathbf E
    \\
    &=-\left(
        \frac{\partial u_{\rm vac}}{\partial t}
        +\nabla\cdot\mathbf S_{\rm vac}
      \right).
\end{aligned}
    \label{eq:retained-complement-power-balance}
\end{equation}
Under these local-storage conditions, the microscopic material response is represented by two macroscopic energy quantities. The \emph{binding energy density} $u_{\rm bind}$ represents recoverable energy in microscopic fields omitted by retention, while $k_{\rm rel}$ is the kinetic-energy density of ordered electron motion relative to the fixed ions. Binding energy thus preserves the energetic contribution of the unresolved fields without reconstructing their microscopic distribution. The term is used only in this sense here, not for the cohesive or dissociation energy of a bound system.

Under these conditions the exchange becomes a storage rate. Positive $\mathbf J_{\rm total}\cdot\mathbf E$ transfers energy from the retained fields into binding and relative kinetic storage; negative power returns energy to the retained fields. If relative kinetic-energy changes are negligible, the retained power determines the binding-energy rate alone. Section~\ref{subsec:binding-energy-material-response} examines the material information needed to determine recoverable binding storage for linear reversible, nonlinear reversible, and dissipative response.

\subsection{Macroscopic Energy Map and Exchange}
\label{subsec:macroscopic-two-side-ledger}

Under the local-storage assumptions of Sec.~\ref{subsec:isolated-dielectric-setup}, the ordered relative charge response is represented by three energy contributions: retained-field energy $u_{\rm vac}$, binding energy $u_{\rm bind}$ representing recoverable unresolved field energy, and the kinetic energy $k_{\rm rel}$ of ordered electron motion relative to the ions.

These electrical contributions complete the thermomechanical map of Sec.~\ref{sec:thermomechanical-receiver}. Table~\ref{tab:energy-architecture-overview} organizes the resulting energies by their microscopic origin and carrier role. Its rows distinguish carrier kinetic energy from electromagnetic field energy; its columns distinguish unordered fluctuations, common motion and deformation, and ordered relative charge configurations and motion.

\begin{table}[t]
\caption{Energy map of the three macroscopic domains within the classical carrier--field model. The columns distinguish macroscopic domains and carrier roles; the energy rows identify their microscopic origin. Thermal and mechanical entries follow Sec.~\ref{sec:thermomechanical-receiver}; recoverable binding storage uses the reversible local-storage assumptions of Sec.~\ref{subsec:isolated-dielectric-setup}.}
\label{tab:energy-architecture-overview}
\small
\renewcommand{\arraystretch}{1.25}
\begin{tabular}{p{0.15\textwidth}p{0.24\textwidth}p{0.24\textwidth}p{0.24\textwidth}}
\toprule
\raggedright Macroscopic domain
& \raggedright Thermodynamics
& \raggedright Continuum mechanics
& \raggedright Macroscopic electrodynamics \tabularnewline
\midrule
\raggedright Carrier role
& \raggedright Unordered fluctuations
& \raggedright Common motion and deformation
& \raggedright Relative charge configurations and motion \tabularnewline
\midrule
\raggedright Carrier kinetic energy
& \raggedright $u_{\rm th}$: randomized carrier energy
& \raggedright $k_{\rm cont}$: continuum kinetic energy
& \raggedright $k_{\rm rel}$: relative-carrier kinetic energy \tabularnewline
\raggedright Microscopic field energy
& \raggedright $u_{\rm rad}$: fluctuating field contribution
& \raggedright $u_{\rm el}$: reversible deformation-related field energy
& \raggedright $u_{\rm vac}$: retained-field energy\newline
  $u_{\rm bind}$: recoverable binding storage \tabularnewline
\botrule
\end{tabular}
\end{table}

The six entries describe the same two microscopic energy stores. Thermal field fluctuations, elastic storage, and recoverable binding storage are all microscopic electromagnetic energy that the retained fields do not carry, represented through different material variables; only $u_{\rm vac}$ is evaluated explicitly from the retained $\mathbf E,\mathbf B$. Thermal, continuum, and relative kinetic energies likewise represent different roles of carrier motion. The thermomechanical entries were identified with the relative charge configuration held fixed, and the electrical entries with the host fixed and thermal exchange suppressed. The dependence of one entry on the variable of another role, such as that of the binding energy on deformation, is the coupling between the domains examined in Sec.~\ref{sec:variational-closure}. Material laws supply this state dependence and coupling.

\smallskip
\noindent\textit{Energy exchange.}
In Eq.~\eqref{eq:retained-complement-power-balance}, mechanical and thermal energy exchange was excluded, so power transferred from the retained fields changed only binding and relative kinetic storage. We now allow common motion and deformation of the host, together with dissipation. The same retained-field power can therefore also supply mechanical work and thermal energy. Its total remains fixed by the retained Poynting balance, while a material model determines its allocation among the three domains:
\begin{equation}
\begin{aligned}
    \mathbf J_{\rm total}\cdot\mathbf E
    &=-\left(
        \frac{\partial u_{\rm vac}}{\partial t}
        +\nabla\cdot\mathbf S_{\rm vac}
      \right)
    \\
    &\simeq
    \underbrace{
        \mathcal P_{\rm unres}
        +\mathcal P_{\rm rel}^{\rm kin}
    }_{\text{electrodynamic}}
    +\underbrace{
        p_{\rm def}
    }_{\text{mechanical}}
    +\underbrace{
        p_{\rm diss}
    }_{\text{thermal}}.
\end{aligned}
    \label{eq:retained-current-receiving-ledger}
\end{equation}

The terms on the right-hand side connect this power to the energy contributions in Table~\ref{tab:energy-architecture-overview}. The electrical powers describe exchange with unresolved field energy and relative carrier motion. Mechanical power $p_{\rm def}$ supplies common host motion and deformation, while $p_{\rm diss}\geq0$ represents direct dissipative transfer to the thermal contributions. These powers account for storage, transport, and transfers within the material description.

Together, Table~\ref{tab:energy-architecture-overview} and Eq.~\eqref{eq:retained-current-receiving-ledger} complete the common energy map: they identify the microscopic origins of its contributions and connect the retained-field exchange to the three macroscopic domains.

\section{The Energy Route to the Load: What the Retained Fields Determine}
\label{sec:variational-closure}

We now examine how the macroscopic electromagnetic domain couples energetically to the continuum-mechanical domain through deformation of the ionic host. For the quasistatic processes considered here, the relevant electrical storage is represented by the retained-field energy $u_{\rm vac}$ and the recoverable binding energy $u_{\rm bind}$. Changes in these energies, together with externally supplied electrical work, determine the mechanical work transferred to the host. We use this connection to establish when the retained fields and available material data determine the mechanical load at the resolution required by the continuum description.

\subsection{From Electrical Energy to Mechanical Load}
\label{subsec:tm-receiving-balances}

Let $U_{\rm vac}$ and $U_{\rm bind}$ denote the total retained-field and recoverable binding energies of the complete conductor--dielectric assembly. For a virtual deformation of the host, energy conservation gives
\begin{equation}
    \delta W_{\rm em}
    \simeq
    \delta W_{\rm ext,elec}
    -\delta\!\left(U_{\rm vac}+U_{\rm bind}\right).
    \label{eq:integrated-electrical-work-balance}
\end{equation}
Here $\delta W_{\rm em}$ is the mechanical work transmitted to the host. The electrical source work $\delta W_{\rm ext,elec}$ depends on the prescribed controls: it vanishes for fixed conductor charges, whereas fixed voltages permit source work as the charges change. For prescribed controls $\Lambda$, we collect the stored electrical energy and source work in a work potential $U_\Lambda$, with $\delta U_\Lambda=\delta(U_{\rm vac}+U_{\rm bind})-\delta W_{\rm ext,elec}$. The mechanical work is therefore $\delta W_{\rm em}\simeq-\delta U_\Lambda$~\cite{McMeekingLandis2005StoredEnergyDeformableDielectrics,McMeekingLandisJimenez2007VirtualWork,Liu2013EnergyFormulationsElectrostaticsContinuumMedia,SuoZhaoGreene2008NonlinearFieldTheory}.

To specify which mechanical load this work measures, let $q$ parameterize an admissible host deformation $\mathbf u(\mathbf X;q)$ under fixed electrical, thermal, and device controls $\Lambda$. We assume zero total constraint virtual work for this variation~\cite[Sec.~2.1]{Maugin1980VirtualPower}. Its virtual displacement is $\delta\mathbf u=\boldsymbol\eta_q\,\delta q$, with $\boldsymbol\eta_q=\partial\mathbf u/\partial q$. The displacement pattern $\boldsymbol\eta_q$ selects the component of loading to be determined. Any dissipative internal variables are held fixed at their current values during this variation; their role is developed in Sec.~\ref{subsec:dissipation-hysteresis}.

For a host occupying $\Omega$, the same virtual work is performed by the transmitted bulk force density $\mathbf f_{\rm em}^{\rm bulk}$ and boundary traction $\mathbf t_{\rm em}$. Writing $\delta W_{\rm em}=Q_{\rm em}^{(q)}\delta q$ identifies the generalized load conjugate to the selected deformation:
\begin{equation}
\begin{aligned}
    Q_{\rm em}^{(q)}
    &\equiv
    \int_{\Omega}
        \mathbf f_{\rm em}^{\rm bulk}
        \cdot\boldsymbol\eta_q\,dV
    +
    \int_{\partial\Omega}
        \mathbf t_{\rm em}
        \cdot\boldsymbol\eta_q\,dA
    \\
    &\simeq
    -\left.\frac{dU_\Lambda}{dq}\right|_\Lambda .
\end{aligned}
    \label{eq:tm-generalized-em-load}
\end{equation}
Equation~\eqref{eq:tm-generalized-em-load} connects the work potential to the load entering continuum mechanics. Whether the available macroscopic information determines this load is a further question. A reduced potential $U_\Lambda^{\rm ret}$, constructed from the retained fields and supplied material data, is sufficient when its negative derivative reproduces $Q_{\rm em}^{(q)}$ within the required tolerance. Microscopic differences left unspecified by that description must therefore leave the selected virtual work unchanged to this accuracy. We refer to this condition as the \emph{sufficiency criterion}. It is tested by comparison: one pair of admissible realizations with different virtual work shows that a description is insufficient (Sec.~\ref{subsec:output-conditioned-closure-tests}), whereas sufficiency is established relative to a stated class of realizations, for instance those covered by a constitutive law with measured deformation dependence.

The retained field solution along the selected deformation determines $\delta U_{\rm vac}$ and, with the source work, the energy that leaves the retained fields during the deformation, $\delta W_{\rm ext,elec}-\delta U_{\rm vac}$. By Eq.~\eqref{eq:integrated-electrical-work-balance}, this is the sum $\delta U_{\rm bind}+\delta W_{\rm em}$. The retained fields fix the sum and not its parts: to obtain the mechanical work, we must know how the recoverable binding energy changes as the host deforms.

\subsection{Binding Energy and Material Response}
\label{subsec:binding-energy-material-response}

\begin{figure}[!tp]
    \centering
    \includegraphics[width=0.79\textwidth]{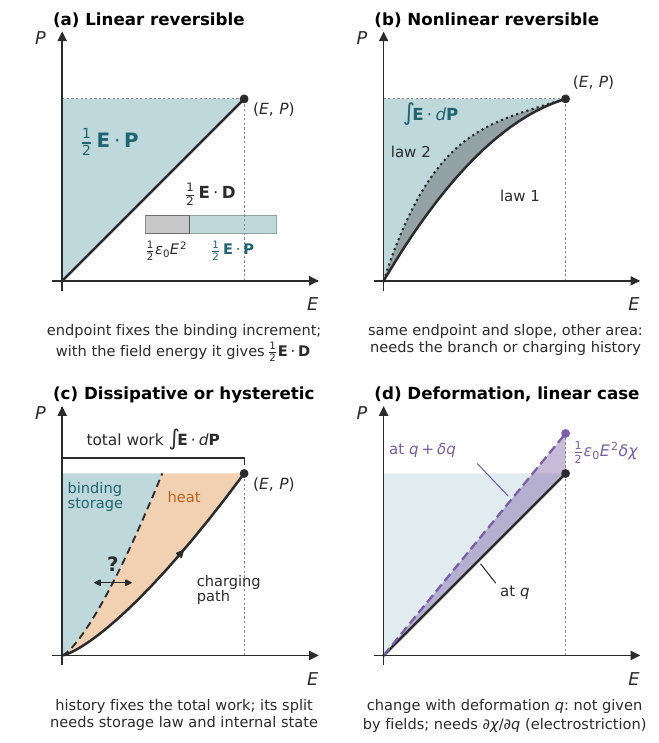}
    \caption{What the retained fields fix about the energy transferred to the material, shown in the $E$--$P$ plane at fixed deformation. The area between a curve and the $P$ axis is $\int\mathbf E\cdot d\mathbf P$; all curves are schematic. (a) Linear reversible response: the endpoint fixes the binding increment $\tfrac12\mathbf E\cdot\mathbf P$, which adds to the retained-field energy $\tfrac12\varepsilon_0|\mathbf E|^2$ to give $\tfrac12\mathbf E\cdot\mathbf D$. (b) Nonlinear reversible response: two laws with the same endpoint and local tangent enclose different areas, so the increment requires the constitutive branch or the charging history. (c) Dissipative or hysteretic response: the charging history fixes the total transferred work; its division into recoverable binding storage and heat, marked by the dashed curve, requires a storage function and the internal state. (d) Deformation at fixed field for linear response: the sliver between the lines at deformations $q$ and $q+\delta q$ has area $\tfrac12\varepsilon_0|\mathbf E|^2\,\delta\chi$, the change of the binding increment, which requires the strain dependence of the susceptibility. For nonlinear and hysteretic response, the corresponding input is the deformation dependence of the branch, or of the storage function at fixed internal state.}
    \label{fig:ep-plane}
\end{figure}

Force prediction requires the recoverable storage at a fixed host configuration and its change with deformation. The material information needed for these two steps depends on the response regime, as Fig.~\ref{fig:ep-plane} illustrates.

At fixed deformation $q$, with $\mathbf J_{\rm free}=\mathbf0$ and $\mathbf M=\mathbf0$, we have $\mathbf J_{\rm total}=\partial_t\mathbf P$. Integrating the retained power balance in Eq.~\eqref{eq:retained-current-receiving-ledger} along the charging history gives the electrical work per unit volume:
\begin{equation}
\begin{aligned}
    -\int\left(
        \frac{\partial u_{\rm vac}}{\partial t}
        +\nabla\cdot\mathbf S_{\rm vac}
    \right)\,dt
    &=\int\mathbf J_{\rm total}\cdot\mathbf E\,dt
    \\
    &=\int\mathbf E\cdot d\mathbf P.
\end{aligned}
    \label{eq:retained-electrical-charging-work}
\end{equation}
The polarization integral thus measures the net energy transferred from the retained fields to the material response, whatever that response is. It is a time-integrated transfer. It becomes the content of a store only when the transfer is recoverable and remains in the material element, and it is in no case the energy of a polarization field: $\mathbf P$ represents the bound current (Sec.~\ref{subsec:retained-total-source-system}) and carries no energy of its own. The familiar density $\tfrac12\mathbf E\cdot\mathbf D$ adds this transfer to the retained-field energy and so presupposes these conditions. Only for linear reversible response does the endpoint alone fix the stored part, and no material responds linearly and reversibly at all field strengths and rates. The regimes below differ in what the transfer history tells about the recoverable part.

\subsubsection{Linear Reversible Response}
\label{subsec:linear-electrostriction}
\label{subsec:linear-binding-closure}

We first consider a linear reversible dielectric. To determine its recoverable binding energy, we charge it from the zero-field reference while holding the host deformation $q$ fixed. Its polarization law is
\begin{equation}
    \mathbf P
    =
    \varepsilon_0
    \boldsymbol\chi(q)\mathbf E
    \label{eq:linear-deformation-dependent-polarization}
\end{equation}
with symmetric $\boldsymbol\chi(q)$, $\mathbf D=\varepsilon_0\boldsymbol\varepsilon(q)\mathbf E$, and $\boldsymbol\varepsilon=\mathbf I+\boldsymbol\chi$.

Under the local-storage conditions of Sec.~\ref{subsec:isolated-dielectric-setup}, with negligible changes in relative kinetic energy, the charging integral in Eq.~\eqref{eq:retained-electrical-charging-work} is the energy transferred into recoverable binding storage. Equation~\eqref{eq:retained-complement-power-balance} therefore gives the binding-energy increment:
\begin{equation}
\begin{aligned}
    u_{\rm bind}(\mathbf E,q)-u_{\rm bind}^{0}(q)
    &=\int_{\mathbf0}^{\mathbf E}\mathbf E'\cdot d\mathbf P(\mathbf E',q)
    \\
    &=\frac{\varepsilon_0}{2}\mathbf E\cdot\boldsymbol\chi(q)\mathbf E
    \\
    &=\frac12\mathbf E\cdot\mathbf P(\mathbf E,q).
\end{aligned}
    \label{eq:linear-deformation-dependent-binding}
\end{equation}
Here $u_{\rm bind}^{0}(q)\equiv u_{\rm bind}(\mathbf0,q)$ is the zero-field reference. The known linear law determines this increment from the retained endpoint [Fig.~\ref{fig:ep-plane}(a)]. This identification assumes a temperature-independent permittivity, consistent with the suppressed thermal exchange of Sec.~\ref{subsec:isolated-dielectric-setup}; if $\boldsymbol\chi$ depends on temperature, the isothermal work $\int\mathbf E\cdot d\mathbf P$ is a change of free energy, and part of it is exchanged as heat with the thermal domain~\cite{LandauLifshitzVol8}. Although it is written in the endpoint fields, $\tfrac12\mathbf E\cdot\mathbf P$ is the value of the transfer integral for this law and not a quantity carried by $\mathbf E$ and $\mathbf P$: the same endpoint reached with a different law corresponds to a different stored part [Fig.~\ref{fig:ep-plane}(b)].

In microscopic terms, Eq.~\eqref{eq:linear-deformation-dependent-binding} states that the binding increment is the change of the residual-field energy: with Eq.~\eqref{eq:microscopic-complementary-field-energy} and negligible cross terms in the smooth bulk, $u_{\rm bind}(\mathbf E,q)-u_{\rm bind}^{0}(q)=\tfrac{\varepsilon_0}{2}\,\Delta\langle|\delta\mathbf e|^2\rangle$ in the electrostatic case, where $\Delta$ denotes the change from the zero-field state. For discrete dipoles, Ref.~\cite{Yaghjian2016ClassicalPowerEnergy} obtains the corresponding relation, in which the polarization work is the work on the dipole carriers plus the change of the energy of the field difference $\mathbf e_{\rm mic}-\mathbf E$; with workless constraints and negligible kinetic energy, the first part stores nothing. For lossless periodic composites, the cell-averaged microscopic energy equals the macroscopic expression formed with the effective dielectric function~\cite{Silveirinha2009PoyntingStoredEnergy}.

Adding this increment to the retained-field energy gives the field-induced recoverable density in the static electric regime:
\begin{equation}
\begin{aligned}
    u_{\rm rec}(\mathbf E,q)
    &=u_{\rm vac}
    +\left[u_{\rm bind}(\mathbf E,q)-u_{\rm bind}^{0}(q)\right]
    \\
    &=\frac12\mathbf E\cdot\mathbf D .
\end{aligned}
    \label{eq:linear-drive-bundled-density}
\end{equation}
The conventional density thus combines the retained-field energy with the recoverable binding-energy increment. Its volume integral supplies the field-induced storage contribution to the work potential $U_\Lambda$. Deformation-dependent contributions from the zero-field reference $u_{\rm bind}^{0}(q)$ and interfaces must also be included when they affect the selected load.

To obtain the mechanical load in Eq.~\eqref{eq:tm-generalized-em-load}, we must determine how the work potential changes with deformation, including changes in the retained fields, geometry, and material response. Knowing the permittivity at one geometry does not determine how the binding energy changes as the host deforms; its deformation dependence is additional material information [Fig.~\ref{fig:ep-plane}(d)].

To express the resulting load as a continuum stress, we take local strain $u_{ik}$ as the deformation variable. For isothermal variations, the work potential is constructed from the Helmholtz free energy $F=U-TS$, with the electrical source and boundary contributions appropriate to the prescribed controls. Its strain variation gives the conventional Helmholtz/Landau--Lifshitz expression for the electrically induced continuum stress in the small-strain limit, for a solid that is isotropic in the undeformed state~\cite{LandauLifshitzVol8,Melcher1981,LeePengShkel2005,Shkel2007,ZhaoSuo2008}:
\begin{equation}
\begin{aligned}
    \sigma_{ik}^{\rm elec}
    &=\frac12\left(E_iD_k+E_kD_i\right)
      -\frac12 E_lD_l\,\delta_{ik}
    \\
    &\quad-\frac{\varepsilon_0}{2}E_lE_m
    \left(
        \frac{\partial\varepsilon_{lm}}{\partial u_{ik}}
    \right)_{T,\mathbf E} .
\end{aligned}
    \label{eq:linear-electrical-continuum-stress}
\end{equation}
Here $u_{ik}$ is the infinitesimal strain and $T$ the temperature. The last, electrostrictive term is the contribution that the retained fields leave open: it arises from the explicit strain dependence, at fixed field, of the binding increment $\tfrac{\varepsilon_0}{2}\mathbf E\cdot\boldsymbol\chi\mathbf E$ [Fig.~\ref{fig:ep-plane}(d)], that is, of energy in field structure that averaging removes. The other terms follow from the retained-field solution and the permittivity at the given configuration. This term has been evaluated for lattices of dielectric cylinders by deforming the unit cell: it depends on the lattice symmetry at fixed effective permittivity, and the Helmholtz stress formed with it is the one that matches the force density computed from the microscopic fields~\cite{SunWangNgZhouChan2015}. A load at one configuration requires the relevant deformation derivative there; prediction over a finite deformation range requires its variation over that range. These material data can be measured or obtained from strained higher-resolution models with the relevant internal relaxation~\cite{Srinivasan1968LatticeElasticDielectric,TannerBousquetJanolin2021Electrostriction}.

\subsubsection{Nonlinear Reversible Response}
\label{subsec:reversible-nonlinear-closure}

We next consider a nonlinear reversible dielectric, for which the charging integral generally does not reduce to $\tfrac12\mathbf E\cdot\mathbf P$. Under the same local-storage assumptions, the recoverable binding-energy increment is determined from a reversible integrable constitutive branch $\mathcal C_{\rm rev}(q)$ connecting the reference and operating states at fixed host deformation $q$:
\begin{equation}
    u_{\rm bind}(\mathbf E,q)
    -
    u_{\rm bind}^{0}(q)
    =
    \int_{\mathcal C_{\rm rev}(q)}
    \mathbf E'\cdot d\mathbf P' .
    \label{eq:nonlinear-deformation-dependent-binding}
\end{equation}
A known retained charging history again determines this increment through Eq.~\eqref{eq:retained-electrical-charging-work}. The retained endpoint alone generally does not: different reversible nonlinear constitutive laws can share the same endpoint $(\mathbf E,\mathbf P,\mathbf D)$ and local tangent while giving different binding increments [Fig.~\ref{fig:ep-plane}(b)]. Determining the increment from endpoint data requires the constitutive branch between the reference and operating states, or equivalent stored-energy data.

\smallskip
\noindent\emph{Deformation dependence.}
Using the branch integral, the field-induced recoverable density and selected load take the form
\begin{equation}
\begin{aligned}
    u_{\rm rec}(\mathbf E,q)
    &=u_{\rm vac}
    +\int_{\mathcal C_{\rm rev}(q)}\mathbf E'\cdot d\mathbf P',
    \\
    Q_{\rm em}^{(q)}
    &=-\left.\frac{dU_\Lambda[u_{\rm rec};q]}{dq}\right|_\Lambda .
\end{aligned}
    \label{eq:nonlinear-force-branch-derivative}
\end{equation}
Here $U_\Lambda[u_{\rm rec};q]$ includes the deformed geometry and all source, boundary, and reference contributions relevant to the selected load. Knowing the branch at one geometry fixes the binding increment there. Force prediction additionally requires how that branch integral changes under the selected deformation.

A deformation-dependent constitutive branch or equivalent stored-energy function supplies this information~\cite{DorfmannOgden2005NonlinearElectroelasticity,BustamanteDorfmannOgden2009,EnergyFormulationMagnetoElectroElasticity2013}. Homogenization provides one route to constructing such a function and its deformation derivatives from a specified microstructure~\cite{PonteCastanedaSiboni2012Homogenization}. Together with the retained-field variation, this information determines the recoverable energy change entering the sufficiency criterion.

\subsubsection{Dissipation and Hysteresis}
\label{subsec:dissipation-hysteresis}

We finally consider a dielectric with a dissipative or hysteretic polarization response. Equation~\eqref{eq:retained-electrical-charging-work} still determines the net energy transferred from the retained fields to the material, but this transfer now supplies both a change in recoverable binding storage and dissipation. The total charging work therefore does not by itself determine the recoverable energy needed for force prediction [Fig.~\ref{fig:ep-plane}(c)]. For preparation at fixed host deformation $q$, with relative-carrier inertia, reversible thermal exchange, and independent transport inactive,
\begin{equation}
\begin{aligned}
    w_{\rm elec}[\mathcal C]
    &\equiv
    \int_{\mathcal C}\mathbf E\cdot d\mathbf P
    =
    \Delta u_{\rm bind}(\mathcal H_0,\mathcal H)
    +
    w_{\rm diss}[\mathcal C],
    \\
    w_{\rm diss}[\mathcal C]
    &\geq
    0 .
\end{aligned}
    \label{eq:path-dependent-host-allocation}
\end{equation}
Here $\mathcal H_0$ and $\mathcal H$ are the initial and current internal states; dependence on the retained endpoint and on $q$ is implicit. The irreversible part $w_{\rm diss}$ feeds the thermal domain of Table~\ref{tab:energy-architecture-overview}; its subsequent storage and transport belong to the thermomechanical description of the host.

The retained electromagnetic history fixes the transferred work in Eq.~\eqref{eq:path-dependent-host-allocation}; a material model supplies its division between recoverable binding storage and dissipation~\cite{MieheRosato2011RateDependentFerroelectricity,FooEtAl2012DissipativeDielectricElastomers}. When the response depends on history, the same instantaneous $(\mathbf E,\mathbf P,\mathbf D)$ can correspond to different internal states $\mathcal H$, with different recoverable energies and deformation derivatives. These internal variables retain the relevant memory of the carrier--field configuration without introducing a new microscopic energy store.

For a dissipative or hysteretic dielectric, the binding storage depends on the current internal state $\mathcal H$. Applying Eq.~\eqref{eq:tm-generalized-em-load} to the corresponding work potential through a virtual deformation with $\mathcal H$ held fixed gives
\begin{equation}
    Q_{\rm em}^{(q)}
    =
    -
    \left.
    \frac{\partial U_\Lambda(q,\mathcal H)}{\partial q}
    \right|_{\Lambda,\mathcal H}
    .
    \label{eq:hysteretic-force-needs-history-closure}
\end{equation}
The retained field solution along the selected deformation determines the retained-field energy variation. The recoverable binding energy and its deformation variation, however, require material information beyond the retained fields: a constitutive storage function and the current internal state. Predicting that state from the retained electromagnetic history additionally requires an initial state and an evolution law~\cite{MieheRosato2011RateDependentFerroelectricity}. Once this information is supplied, the same virtual-work relation determines the conservative electrical driving load.

\subsection{Attainable and Required Force Resolution}
\label{subsec:tm-output-resolution}

We now examine how finely the bulk force density and boundary traction can be resolved from the retained fields and the material information of Sec.~\ref{subsec:binding-energy-material-response}. The result has two parts: each virtual displacement yields one component of the load, so resolving its distribution requires correspondingly varied deformations, and the attainable resolution is limited by that of the retained fields and the supplied data.

The required force resolution depends on the mechanical prediction of interest. Predicting deformation at a specified spatial resolution often requires only the collective loading of many particles or dipoles. Their individual force peaks need not be resolved, provided the averaged description preserves the selected mechanical response to the required accuracy.

\begin{figure}[tbp]
    \centering
    \includegraphics[width=\textwidth]{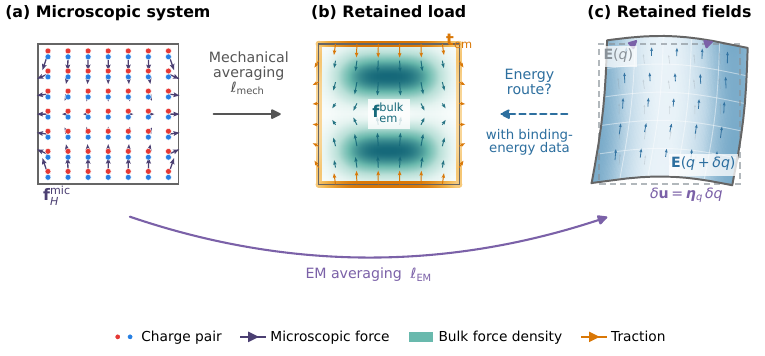}
    \caption[Two routes to the retained mechanical load.]{Two routes to the retained mechanical load. (a) An illustrative $7\times7$ point-dipole array and its host forces, with fourth-root arrow-length scaling. Gray contours mark the host boundary. Mechanical retention, (a)$\to$(b), represents interior forces as $\mathbf f_{\rm em}^{\rm bulk}$ and the boundary layer as traction $\mathbf t_{\rm em}$. Electromagnetic retention, (a)$\to$(c), averages the microscopic fields of the same configuration. Panel (c) schematically shows the retained fields under virtual deformation. The dashed route (c)$\to$(b) tests agreement in virtual work [Eq.~\eqref{eq:tm-generalized-em-load}], using binding-energy data in addition to the retained fields.}
    \label{fig:mechanical-load-retention}
\end{figure}

We represent this collective loading by a bulk force density and boundary traction. In Fig.~\ref{fig:mechanical-load-retention}, the direct route (a)$\to$(b) obtains these quantities from the microscopic host load $\mathbf f_H^{\rm mic}$, with carrier-to-host transmission specified:
\begin{equation}
\begin{aligned}
    \mathbf f_{\rm em}^{\rm bulk}
    &\simeq
    \left\langle
        \mathbf f_H^{\rm mic}
    \right\rangle_{\ell_{\rm mech}},
    \qquad
    d_{\rm mic}
    \ll
    \ell_{\rm mech}
    \ll
    L_{\rm out},
    \\
    \mathbf t_{\rm em}
    &\simeq
    \int_{\rm transition\ layer}
        \mathbf f_H^{\rm mic}\,dn .
\end{aligned}
    \label{eq:tm-mechanical-load-retention}
\end{equation}
Here $L_{\rm out}$ is the smallest spatial scale the mechanical prediction must resolve, and $n$ is distance normal to the interface. The bulk average suppresses microscopic force variations while retaining their collective loading. At an unresolved interface, integration through the force layer retains its action as a boundary traction, with tangential averaging at the chosen mechanical resolution. These bulk and boundary loads define the continuum target for the subsequent energy-based prediction. Carrier-to-host transmission and mechanical retention are developed in Ref.~\cite[Secs.~III--IV]{SchulteWesthoffForceIndeterminacy2026}.

For prediction, we use the retained fields at scale $\ell_{\rm EM}$, obtained by electromagnetic retention along (a)$\to$(c), supplemented by the binding-energy information identified in Sec.~\ref{subsec:binding-energy-material-response}. These inputs enter the work potential of Sec.~\ref{subsec:tm-receiving-balances}. Along the dashed route (c)$\to$(b), we examine at which mechanical resolutions its variations reproduce the work of the directly averaged bulk and boundary loads.

Equation~\eqref{eq:tm-generalized-em-load} relates the force information obtained to the chosen virtual displacement. A rigid translation determines the total force in its direction. Resolving the load distribution requires energy changes under spatially varying deformations at the required mechanical resolution. Individual particle or dipole forces can remain unresolved when their collective work under these deformations is captured.

For the smooth-bulk description considered here, the retained fields must resolve the spatial variations needed for the selected mechanical load. When no finer spatial information is supplied separately, this requires
\begin{equation}
    \ell_{\rm EM}\lesssim\ell_{\rm mech}.
    \label{eq:force-resolution-condition}
\end{equation}
At this resolution, material and interface information must also determine the changes of unresolved field energy contributing to the selected work. A localized defect, for example, may alter the binding-energy variation under deformation while remaining unresolved in the retained electromagnetic response. If significant at the required accuracy, this contribution must be captured by material data or a locally resolved microscopic model. Atomistic-to-continuum methods with atomic resolution near defects provide one such route~\cite{JhaEtAl2023DefectsElectrostatics}.

\subsection{Identical Retained Field Paths, Different Binding-Energy Variations}
\label{subsec:output-conditioned-closure-tests}

We examine the energy consequence of the aligned--staggered boundary comparison established in Ref.~\cite{SchulteWesthoffForceIndeterminacy2026} (Sec.~V\,C and Appendix~A). Two arrays of identical parallel line dipoles have the same number density, continuum body, and carrier-to-host transmission. In the aligned arrangement $A$, neighboring rows are vertically registered; in the staggered arrangement $S$, successive rows are offset laterally by half a lattice spacing.

This offset lies below the retained spatial resolution, leaving the bulk polarization unchanged. With the same sharp-interface polarization profile and electrical boundary conditions, both arrays therefore have the same retained electric field. Their retained bulk loads agree, but boundary-layer integration and surface averaging give different tractions, $\mathbf t_A\ne\mathbf t_S$. At the top boundary, staggering replaces a directly underlying neighbor by diagonal neighbors, changing the near-field gradients sampled by each dipole. In the line-dipole benchmark of Ref.~\cite{SchulteWesthoffForceIndeterminacy2026} (its Fig.~2), the top traction, normalized by the field-only value common to both arrangements, is $-0.149$ for the aligned and $+0.145$ for the staggered arrangement, against $+1.00$ for the field-only expression itself. These values characterize that benchmark; the conclusion used here is that the tractions differ at identical retained data.

For the energy comparison, we additionally prescribe identical retained polarization profiles throughout a common quasistatic virtual deformation, parameterized by $q$, with the same sharp-interface convention and electrical boundary conditions. This common deformation response is an additional assumption of the present example. It gives $\mathbf E_A(q)=\mathbf E_S(q)$ and hence $\delta U_{{\rm vac},A}=\delta U_{{\rm vac},S}$; the common fields may vary with $q$. Equality of these paths does not fix the unresolved near fields or their changes during deformation.

We choose a smooth continuum virtual displacement $\delta\mathbf u$ for which the two boundary loads perform different virtual work at the reference configuration. The equal bulk loads and retained-field energy variations cancel in the comparison. Under the conservative virtual-work assumptions of Sec.~\ref{subsec:tm-receiving-balances}, with rigid fixed-moment dipoles, workless internal constraints, and no external electrical input, energy conservation gives
\begin{equation}
\begin{aligned}
    \delta W_S-\delta W_A
    &=\int_{\partial\Omega}
    (\mathbf t_S-\mathbf t_A)\cdot\delta\mathbf u\,ds\\
    &=-\delta\!\left(U_{{\rm bind},S}-U_{{\rm bind},A}\right)
    \ne0.
\end{aligned}
\label{eq:output-test-aligned-staggered-hidden-binding-derivative}
\end{equation}
Here $\Omega$ is the planar cross-section; work and energy are per unit out-of-plane length. The dipole self-energies remain unchanged, so the work difference comes from different variations of mutual interactions in the unresolved field energy, which includes retained--residual cross terms at unresolved interfaces [Eq.~\eqref{eq:microscopic-complementary-field-energy}].

Under the prescribed common deformation response, the unequal boundary tractions therefore correspond to different binding-energy variations. The retained-field path alone leaves these variations unspecified. Completing the route (c)$\to$(b) in Fig.~\ref{fig:mechanical-load-retention} requires material or interface information that determines the binding-energy variation contributing to the selected mechanical work.

\subsection{Implications for Electromechanical Modeling}
\label{subsec:constructive-closure-routes}

The sufficiency criterion gives a practical rule for coupling a retained electromagnetic description to continuum mechanics: the fields and supplied material data must together determine the work of the loads required by the mechanical model. Begin with the intended deformation or load distribution, including any relevant boundary tractions, and specify the spatial resolution, accuracy, and prescribed controls. These choices identify the virtual deformations whose work the description must reproduce and the spatial information required by Sec.~\ref{subsec:tm-output-resolution}.

The energetic input is the variation of retained-field and recoverable binding energies, together with the source work appropriate to the controls. The material information identified in Sec.~\ref{subsec:binding-energy-material-response} must determine the binding-energy variation for the selected deformations, including relevant reference and interface contributions. Recoverable storage must be distinguished from losses, with internal-state and evolution information supplied where needed. Table~\ref{tab:retained-fields-determine} collects, for each quantity that enters this route, what the retained fields determine and what must be supplied in addition. When these inputs are known, Eq.~\eqref{eq:tm-generalized-em-load} gives the selected electrical driving load under the virtual-work assumptions of Sec.~\ref{subsec:tm-receiving-balances}.

\begin{table}[!htb]
\caption{What the retained fields determine about the quantities used by the energy route. Each statement holds under the assumptions of the section cited.}
\label{tab:retained-fields-determine}
\small
\renewcommand{\arraystretch}{1.25}
\begin{tabular}{@{}p{0.26\textwidth}p{0.25\textwidth}p{0.31\textwidth}p{0.08\textwidth}@{}}
\toprule
\raggedright Quantity
& \raggedright Determined by the retained fields?
& \raggedright Additional information required
& Section \tabularnewline
\midrule
\raggedright Energy and energy flux of the retained fields, $u_{\rm vac}$ and $\mathbf S_{\rm vac}$
& \raggedright Yes
& \raggedright None
& \ref{subsec:external-power-conductor-charging} \tabularnewline
\raggedright Work supplied by the external source
& \raggedright Yes, up to surface-layer corrections
& \raggedright None
& \ref{subsec:external-power-conductor-charging}, \ref{subsec:isolated-dielectric-setup} \tabularnewline
\raggedright Net power exchanged with the material, $\mathbf J_{\rm total}\cdot\mathbf E$
& \raggedright Yes
& \raggedright Material laws for its division among storage, motion, deformation, and heat
& \ref{subsec:isolated-dielectric-setup}, \ref{subsec:macroscopic-two-side-ledger} \tabularnewline
\raggedright Recoverable binding energy at fixed deformation, linear reversible response
& \raggedright Yes, with the linear law: increment $\tfrac12\mathbf E\cdot\mathbf P$
& \raggedright None beyond the linear law
& \ref{subsec:linear-binding-closure} \tabularnewline
\raggedright The same, nonlinear reversible response
& \raggedright Only with the charging history, not from the endpoint
& \raggedright Constitutive branch or stored-energy function
& \ref{subsec:reversible-nonlinear-closure} \tabularnewline
\raggedright The same, dissipative or hysteretic response
& \raggedright No; only the total transferred work
& \raggedright Storage function and internal state; an evolution law to predict that state
& \ref{subsec:dissipation-hysteresis} \tabularnewline
\raggedright Change of the binding energy under deformation, which enters the load
& \raggedright No
& \raggedright Deformation dependence of the permittivity, constitutive branch, or storage function; interface or spatial information where unresolved structure changes the virtual work of interest
& \ref{subsec:binding-energy-material-response}--\ref{subsec:output-conditioned-closure-tests} \tabularnewline
\botrule
\end{tabular}
\end{table}

Missing deformation dependence can often be supplied at the existing continuum resolution through constitutive data or a homogenized stored-energy function~\cite{PonteCastanedaSiboni2012Homogenization}. Where omitted structure changes the selected work beyond the required accuracy, targeted microscopic or multiscale calculations can supply the relevant energy variations or forces~\cite{ParkDevelWang2011MultiscaleElectromechanics,MarshallDayal2014}. A validated force or stress law, or an experimentally calibrated load, also provides a direct route; Ref.~\cite{SchulteWesthoffForceIndeterminacy2026} illustrates boundary-load reconstruction. For the same conservative realization and controls, the resulting energy and force descriptions must agree in virtual work.

\section{Conclusion}
\label{sec:conclusion}

The energy route to the mechanical load on a dielectric requires the stored energy and its change under deformation. We have examined how much of this the retained fields determine (Table~\ref{tab:retained-fields-determine}). The retained fields carry their own energy, and their balance has one exchange term, $\mathbf J_{\rm total}\cdot\mathbf E$: the power leaving them. Up to surface-layer corrections they determine the external work supplied to the assembly, and they determine this exchange; they do not show where it goes. The polarization integral $\int\mathbf E\cdot d\mathbf P$ is accordingly a time-integrated transfer out of the retained fields and not the energy of a polarization field. The state of the retained fields at an instant thus fixes only their own energy, and their history fixes the energy transferred; what remains stored, and how it changes under deformation, is material information. Under deformation the retained fields therefore fix the sum of binding-energy change and mechanical work, and predicting a selected load requires the relevant deformation dependence of the energy.

What lies beyond the exchange follows from the microscopic model. The exchange is not the averaged work on the carriers: part of it feeds field structure that averaging removes. Under the stated local-storage conditions, recoverable energy of this structure is represented by binding energy, and in the linear reversible electrostatic regime $\tfrac12\mathbf E\cdot\mathbf D$ combines retained-field energy with the binding increment $\tfrac12\mathbf E\cdot\mathbf P$. Knowing this storage per unit volume does not by itself specify its microscopic spatial distribution or its deformation dependence. Within the classical model, carrier kinetic energy and microscopic electromagnetic field energy are the two active stores; averaging and collective variables represent their contributions through thermal, mechanical, and electrical energies. The resulting energy map identifies the physical content shared by thermodynamics, continuum mechanics, and macroscopic electrodynamics and the exchanges connecting them.

The available description is sufficient when all microscopic realizations consistent with it give the same virtual work to the required accuracy. The energy route then yields the load component by component, one for each virtual deformation, and no finer than the retained fields and the supplied data resolve it. The aligned--staggered comparison, extended by assuming identical retained fields throughout the deformation, illustrates that identical retained field paths can accompany different binding-energy variations and continuum boundary work. The required material information can often be supplied at the existing continuum resolution. Deformation-dependent constitutive laws, recoverable storage functions, and internal-state descriptions can encode the collective mechanical effect of unresolved fields without reconstructing their microscopic distribution. Additional spatial or interface information is needed when omitted structure changes the selected work beyond the required accuracy.

The energy route therefore does not remove the need for information beyond the macroscopic fields that appears on the force route~\cite{SchulteWesthoffForceIndeterminacy2026}: for the two dipole arrangements, the difference in boundary work is given by the difference of their binding-energy variations. The two routes meet the same structure. Because the retained fields obey Maxwell's equations, the identities of Sec.~\ref{subsec:carrier-ownership-ledger} hold for them as well: they exchange energy through $\mathbf J_{\rm total}\cdot\mathbf E$ and momentum through the Lorentz force on the retained sources, and in neither case do they show where the exchange goes. The common energy map establishes the physical basis for coupling the macroscopic domains, while the sufficiency criterion specifies when their available material descriptions determine the mechanical load. The present analysis is classical and restricted to electric response; intrinsic spin and magnetic matter are left for future work. The same construction can be extended from energy to momentum, stress, and transport, which bears on the identification of the electromagnetic energy--momentum tensor in matter; this extension is also left to separate work.

\backmatter

\section*{Statements and Declarations}

\begin{itemize}
\item \textbf{Competing interests} The author has no competing interests to declare that are relevant to the content of this article.
\item \textbf{Funding} No funding was received for conducting this study.
\item \textbf{Data and code availability} The Python code used to generate Figs.~\ref{fig:source-field-retention-comparison} and~\ref{fig:mechanical-load-retention} is available from the author on reasonable request. Figure~\ref{fig:ep-plane} is a schematic of analytic curves and involves no simulation data. The theoretical analysis is presented in this article; the aligned--staggered boundary-traction result is established in Ref.~\cite{SchulteWesthoffForceIndeterminacy2026}.
\item \textbf{Use of generative AI tools} The author used large language models from OpenAI (GPT) and Anthropic (Claude) for assistance in drafting and revising the text, in writing the Python code that produces the figures, and in checking calculations. The author verified the scientific content by working through the derivations and comparing with the literature, and takes full responsibility for the article.
\end{itemize}

\bibliography{Paper_energy_variationalReferences}

\end{document}